\documentclass[12pt]{article} 
\usepackage[margin=1in]{geometry}
\usepackage{lineno}
\usepackage{physics}
\usepackage{hyperref}
\usepackage{amsmath}
\usepackage{setspace}
\usepackage{physics}
\usepackage{slashed}
\usepackage{amssymb}
\usepackage{color}
\usepackage{url}
\usepackage{epstopdf}
\usepackage{authblk}
\usepackage{siunitx}
\usepackage{blochsphere}
\usepackage{graphicx}
\usepackage{wrapfig}
\usepackage[labelfont=bf]{caption}
\usepackage{tikz}
\usepackage{color}
\usetikzlibrary{shapes,decorations,arrows,calc,arrows.meta,fit,positioning}
\tikzset{
	-Latex,auto,node distance =1 cm and 1 cm,semithick,
	state/.style ={ellipse, draw, minimum width = 0.7 cm},
	point/.style = {circle, draw, inner sep=0.04cm,fill,node contents={}},
	bidirected/.style={Latex-Latex,dashed},
	el/.style = {inner sep=2pt, align=left, sloped}
}

\usepackage[authoryear,round]{natbib}
\newcommand{\indep}{\perp \!\!\! \perp}

\title{Entanglement, Complexity, and Causal Asymmetry in Quantum Theories}
\author{Porter Williams\\
	Department of Philosophy\\
	University of Southern California}
\date{}                     %% if you don't need date to appear
\begin{document}
	\maketitle
	\onehalfspacing
	%I\linenumbers
	
\begin{abstract}
\noindent It is often claimed that one cannot locate a notion of causation in fundamental physical theories. The reason most commonly given is that the dynamics of those theories do not support any distinction between the past and the future, and this vitiates any attempt to locate a notion of causal asymmetry --- and thus of causation -- in fundamental physical theories. I argue that this is incorrect: the ubiquitous generation of entanglement between quantum systems grounds a relevant asymmetry in the dynamical evolution of quantum systems. I show that by exploiting a connection between the amount of entanglement in a quantum state and the algorithmic complexity of that state, one can use recently developed tools for causal inference to identify a causal asymmetry -- and a notion of causation -- in the dynamical evolution of quantum systems.
	\end{abstract}

\section{Introduction}

Consider the plight of Emma Flake. Dr.\ Flake runs a lab that works on applications of quantum state tomography, but she has kind of checked out lately.
%\footnote{Quantum state tomography will play no explicit role in this paper, but see \citep{2004_altepeter} or \citep{2009_lvovsky_raymer}.} 
Her grad students,  Alice and Bob, have conducted the following experiment in her absence: a source prepares an ensemble of bipartite quantum systems in identical pure states $\ket{\alpha}$. Alice knows the state $\ket{\alpha}$, but Bob does not. Each system then passes through a region of spacetime governed by a Hamiltonian $H$, ending up in some pure state $\ket{\beta}$. Bob's task is to reconstruct the pure state $\ket{\beta}$ by performing a large number of  measurements on different observables of the system.

Dr.\ Flake, feeling guilty about her absenteeism, offers to write up the paper on her own and tells Alice and Bob to take some time off. When she looks at Alice and Bob's notes, she discovers that there is no record of which state was prepared by Alice and which was reconstructed by Bob; all she knows is the two pure states $\ket{\alpha}$ and $\ket{\beta}$. Is there any way for Dr.\ Flake to determine which was the initial state and which was the final? In other words, can Dr.\ Flake determine whether $\ket{\alpha}$ was caused by time-evolving $\ket{\beta}$ or vice versa?\footnote{One might wonder why I describe this as a problem of inferring causal direction rather than temporal direction. I will return to this question in section 6.}

Dr. Flake is confronted with a special case of the general problem I will take up in this paper: if one knows (i) the identity of two quantum states and (ii) that they are related by unitary time evolution, does quantum theory provide resources to determine which state was caused by time-evolving the other? It is commonly said that the answer  is ``no'': the failure of the dynamical equations describing microscopic physical systems to distinguish between past and future leaves Dr.\ Flake with insufficient resources to solve her problem. This failure is commonly presented as an obstacle to identifying any notion \textit{at all} of causal direction -- and thus of causation --  using the resources of microscopic physics (see, for example, \citep{1912_russell,2000_albert,2003_field,2007_woodward,2012_loewer}).

One consequence has been the widespread belief that any notion of causal direction, and thus causation, must be located elsewhere. For many, like \citep{2000_albert,2007_woodward}, that has been in macroscopic systems. For some, it has been in our psychological experience as deliberators \citep{2016_ismael,2017_fernandes}. And for a few, it has been nowhere: causal direction is illusory, tantamount to a preferred choice of coordinates \citep{2007_price}.%\footnote{An exception to the belief that the failure of the laws governing microsopic systems to didstinguish past and future undermines any attempt to locate a microscopic direction of causation is \citep{2007_maudlin}. Maudlin argues that the direction of time is ontologically primitive and determines the direction of causation.}

The notion of causation I employ in this paper is a minimal interventionist one: X is a cause of Y if and only if there is an intervention that can be performed on the value of X, while holding all other variables fixed, that produces a change in the value of Y.\footnote{This generalizes straightforwardly to probabilistic settings: X causes Y if and only if there is an intervention that can be performed on $p_X(x)$, the probability distribution over the values of X, while holding all other variables fixed, that produces a change in $p_Y(y)$. See \cite{2005_woodward} for a much richer philosophical development of the interventionist account of causation than I will need here.} In this paper I will restrict to the case of two variables and presuming the absence of confounders, so any causal relation will be a \textit{direct} causal relation. %Quantum states define probability distributions over the possible measured values \textit{y} of observables Y so this probabilistic generalization will be particularly useful for my purposes. 

It is well known that without additional conditions, this minimal interventionist notion of causation cannot use observational statistical data alone to identify a direction of causation between two variables X and Y, even when they are related by a known invertible function (see, e.g., \citep[chapter 2]{2009_pearl}). Unitary time evolution between two quantum states is deterministic and invertible; without information about the actual time ordering of two states $\ket{\alpha}$ and $\ket{\beta}$, familiar causal inference methods are unable to distinguish between $\ket{\alpha} \longrightarrow  \ket{\beta}$ and $\ket{\beta} \longrightarrow  \ket{\alpha}$.

In this paper, I argue that a generic physical fact about the time evolution of quantum systems identifies a causally relevant asymmetry: interaction between quantum systems almost always entangles those systems but almost never disentangles them. This fact enables one to identify a causally relevant asymmetry between quantum states related by unitary time evolution. To connect this physical fact to formal methods of causal inference, I employ recently developed tools whose applications include the ability to identify the causal direction of deterministic, noiseless processes solely from observational data \citep{2008_janzing_scholkopf,2010_daniusis_janzing_mooij,2012_daniusis_janzing,2017_peters_janzing_scholkopf}, \citep[sections 3 and 5.2]{2016_mooji_peters_janzing} to argue that the generic entanglement of quantum systems by unitary time evolution allows one to identify criteria for inferring causal direction even between states related by time-symmetric dynamical laws. The majority of my discussion focuses on paradigmatically microscopic physical systems: bipartite quantum systems undergoing unitary time evolution. That one can identify causally relevant asymmetries in such evolutions is indication that one can locate a notion of causation in microscopic physics after all.

The structure of the paper is as follows. In Section 2, I review the sense(s) in which quantum theories fail to distinguish between past and future and the challenge this poses to any attempt to locate a notion of causation in microscopic physics. In Section 3, I review relevant properties of entanglement in quantum theories: the ubiquity of entangled states, the generic creation of entanglement by interactions between systems, and quantitative measures of how entangled a quantum state is. I argue that already, on physical grounds, these features ground a (fairly weak) causal asymmetry between quantum states related by unitary evolution. Section 4 reviews the causal inference methods I later use to argue that one can identify a direction of causation even between states related by time-symmetric dynamical laws, along with some necessary concepts from algorithmic information theory. Section 5 introduces a generalization of algorithmic information to quantum theories  and demonstrates that in the restricted but important setting of a system of \textit{N} qubits, one can derive a stronger version of the asymmetry from section 3 as a theorem of the causal inference methods introduced in section 4. In section 6 I conclude with some remarks about the relationship between more familiar temporal asymmetries in physics and the causal asymmetry argued for here, and situate my discussion within a more general understanding of the relationship between the epistemology of causation and its metaphysics.

I want to emphasize three things before proceeding.\footnote{Well, four things. The fourth thing is that throughout the paper $\log = \log_2$ and I adopt units where $\hbar=c=1$.} The first thing is that my primary interest here is the \textit{epistemology} of causal direction.\footnote{See \citep{2009_eberhardt} for an overview. For a clarifying sketch of several types of projects one might pursue when thinking about causation, see \citep{2014_woodward}.} One can view sections 4 and 5 in particular as an exercise in causal epistemology: taking a procedure for inferring causal direction based on observational data that has proved successful in other domains and applying it to quantum theories. I think of this as akin to expanding an experimental technique that has proved successful at detecting a certain object or property in one department of nature to a novel domain, where the presence of that object or property is not settled, and seeing what  the experiment turns up. In section 6 I return to a discussion of how this primarily epistemological exercise bears on questions about the physical ground of the causal asymmetry I identify. 

The second thing is that unless otherwise stated, I will be considering only unitary time evolution. Once departures from unitary evolution are allowed, either through the tracing out of the environment associated with decoherence \citep[chapter 9]{2012_wallace}, spontaneous collapse of GRW or CSL type \citep[chapter 7]{2000_albert}, or old-fashioned measurement-induced collapse, then the time evolution distinguishes between past and future and an asymmetry between cause and effect is introduced.\footnote{This doesn't mean that all causal inference puzzles are solved once one allows departures from unitary evolution. Far from it: it is only after considering such departures that one encounters the most vexing causal inference problem posed by quantum theories: the explanation of EPR-type correlations.}

The third thing concerns the discussion of entanglement. Entanglement of quantum systems by interaction is a generic physical fact. In this paper I employ a particular measure of entanglement, the Schmidt measure, for reasons discussed in section 3. The reader should bear in mind that this is only one of multiple ways to quantify entanglement and the physical fact that interactions generically entangle but almost never disentangle -- the physical foundation for the causal inference strategy developed in this paper -- does not depend on this particular choice of entanglement measure nor the particular causal inference methods used. Indeed, the idea that the generation of entanglement by dynamical evolution is connected to a closely related asymmetry, the asymmetry of time, has been explored by a number of authors \citep{2006_popescu_short_winter,2008_reimann,2009_linden_popescu_short_winter,2010_jennings_rudolph,2012_short_farrelly,2013_goldstein_hara_tasaki,2014_malabarba_et_al,2015_goldstein_hara_tasaki}.

Finally, this paper contributes to a large and rapidly growing literature on causal structure and causal inference in quantum theories \citep{2013_leifer_spekkens,2019_barrett_lorenz_oreshkov,2009_chiribella_et_al,2012_oreshkov_et_al,2017_brukner_et_al,2016_costa_shrapnel,2015_ried_janzing_spekkens_et_al,2017_barrett_spekkens_et_al,2015_wood_spekkens,2015_chaves_majenz_gross,2014_chaves_luft_gross}. A number of mathematical and conceptual frameworks for analyzing causal structure have been developed in this literature. For example, \citep{2013_leifer_spekkens} develop a generalization of Bayesian inference and an associated concept of a conditional quantum state, while the process matrix formalism \citep{2012_oreshkov_et_al,2019_barrett_lorenz_oreshkov} introduces the ``process matrix'' which generalizes the standard notions of a quantum state and a quantum channel and has proven especially valuable for analyzing quantum processes with indeterminate causal order. And some of this work has taken up a similar project as I undertake here: locating causal or temporal asymmetries in the structure of quantum theory \citep{2018_thompson_et_al,2020_schmid_selby_spekkens,2021_hardy,2021_rovelli_et_al}. In the course of this paper, I introduce an additional set of tools into this literature by taking the initial steps toward generalizing the algorithmic-information-theoretic framework for causal inference developed by \citep{2008_janzing_scholkopf} to quantum theories.

\section{Time Evolution in Quantum Theories}

The time evolution of a quantum system over an interval of duration $t$ is described by the unitary operator U(t) = $\displaystyle e^{-iHt}$ where $H$ is the Hamiltonian governing the system.\footnote{This presumes time-translation invariance. Time-translation invariance will always be assumed in this paper, a reflection of the fact that I am considering closed quantum systems.} These unitary operators form a group: Stone's theorem ensures that as long as the Hamiltonian $H$ is self-adjoint then it defines a strongly continous one-parameter group of unitary operators U(t) (see, e.g. \citep[chapter 10.2]{2013_hall}). These unitary operators describe time evolution.

Quantum theories can fail to distinguish between past and future in each of the following two senses:\footnote{See \citep[section 4]{2002_earman} or \citep{2013_farr_reutlinger} for a discussion of the importance of distinguishing between the two when considering how the time-symmetry of a theory bears on our ability to distinguish cause and effect in microphysics.}

\begin{enumerate}
	\item Given a state $\ket{\psi(t)}$, the unitary dynamics determine both the earlier state $\ket{\psi(t - \tau)}$ and the later state $\ket{\psi(t + \tau)}$ for arbitary $\tau$. 
\end{enumerate}

\noindent Quantum theories \textit{always} fail to distinguish between past and future in this sense. It is true whether or not the theory is time-reversal invariant. It is a consequence of the fundamental fact secured by Stone's theorem: the time evolution operators U(t) form a one-parameter group, which ensures that every U(t) has an inverse U(t)$^{-1}$.  Given a quantum state $\ket{\psi(t=0)}$, the operator U(t)= $\displaystyle e^{-iHt}$ uniquely determines the future state $\ket{\psi(t)}$ and its inverse U(t)$^{-1} = \displaystyle  e^{iHt}$ uniquely determines the past state $\ket{\psi(-t)}$. A consequence of this is that given any two quantum states related by unitary time evolution, we have $\ket{\alpha} = \text{U(t)}\ket{\beta}$ and $\ket{\beta} = \text{U(t)}^{-1}\ket{\alpha}$. If one does not have any additional information about their temporal ordering then one cannot determine whether $\ket{\alpha}$ causes $\ket{\beta}$ or vice-versa.\footnote{The distinct but related question of whether the two unitaries U(t)and U(t)$^{-1}$ are equally implementable in practice is complicated \citep{2002_janzing_wocjan_beth,2006_janzing,2018_janzing_wocjan}.}%\footnote{Interesting things happen to time evolution if the Hamiltonian is not self-adjoint. For example, an operator O is \textit{maximal symmetric} if it satisfies O\textbf{v}=O$^\dagger$\textbf{v} for all vectors \textbf{v} in its domain and has no self-adjoint extension. If the Hamiltonian is allowed to be maximal symmetric, an analogue of Stone's theorem ensures that the time evolution operators U(t) form a strongly continuous one-parameter \textit{semigroup}. The existence of inverses U(t)$^{-1}$ is not guaranteed, with the result that time-evolution is uniquely defined for $t \geq 0$  or $t \leq 0$ but not both. See \citep{2018_roberts} for discussion of this and other interesting consequences of relaxing the requirement that observables in quantum theories be self-adjoint.}

It is worth pausing here to address a natural question: isn't it obvious that whichever evolution is generated by U(t) is the true one, with U(t)$^{-1}$ generating the backward-in-time, or acausal, evolution? The reason this does not work is that without antecedent information about the direction of time, one is really considering \textit{two} candidate time variables: \textit{t} and \textit{T}, related by $T = -t$. Whether one considers U(t) as generating forward-in-time evolution or backward-in-time evolution depends on the time variable one chooses. The notation U(t)$^{-1}$ is thus a bit misleading; writing the same unitary operator as U(T) makes the symmetry between \textit{t} and \textit{T} manifest.

To illustrate this by the example above, if one chooses \textit{t} as the time variable then given an initial state $\ket{\psi(t=0)}$, the operator U(t)= $\displaystyle e^{-iHt}$ uniquely determines the future state $\ket{\psi(t)}$ and its inverse U(t)$^{-1} = \displaystyle  e^{iHt}$ uniquely determines the past state $\ket{\psi(-t)}$. Using \textit{T}, the temporal ordering is inverted: $\displaystyle e^{-iHT}$ uniquely determines the \textit{past} state $\ket{\psi(-T)}$ and its inverse U(T)$^{-1} = \displaystyle  e^{iHT}$ uniquely determines the \textit{future} state $\ket{\psi(T)}$. Any attempt to determine the causal direction that requires first choosing between \textit{t} or \textit{T} would beg the question. (See \citep{2019_donoghue_menezes,2020_donoghue_menezes} for a pedagogical presentation and additional discussion of the facts above. In particular, see their interesting discussion of the fact that a choice of the time variable is fixed by the choice of a sign convention for the canonical commutation relations $\left[x, \, p\right] = \pm i\hbar$.)

In addition to the invertibility of the dynamics, a quantum theory may also be time-reversal invariant:\footnote{The appropriate understanding of time-reversal in quantum mechanics has received a fair amount of philosophical attention in recent years \citep{2000_albert,2000_callender,2002_earman,2017_roberts,2019_allori,2020_farr,2019_donoghue_menezes,2020_donoghue_menezes,2020_callender,2020_struyve}. The initial stimulation for much of this work were the arguments for a non-standard definition of time reversal by \citep{2000_albert} and \citep{2000_callender}. For reasons compactly summarized in \citep{2019_roberts}, I remain partial to the traditional account and will adopt it throughout this paper.}

\begin{enumerate}\setcounter{enumi}{1}
	\item There exists a time-reversal operator $\mathcal{T}$ such that 
	
	\begin{enumerate}
		\item $\mathcal{T}$ commutes with the Hamiltonian of the theory, $[\mathcal{T}, \, H] = 0$, and 
		\item If the state $\ket{\psi_0(t=0)}$ evolves under \textit{H} as $\ket{\psi_0(t=0)}$,  $\ket{\psi_1(t_1)}$, $\ldots$, $\ket{\psi_n(t_n)}$, then the state $\mathcal{T}\ket{\psi_n(-t_n)}$ evolves under $\mathcal{T}H\mathcal{T}^{-1}=H$ as $\mathcal{T}\ket{\psi_n(-t_n)}$,  $\mathcal{T}\ket{\psi_{n-1}(-t_{n-1})}$ $\ldots$, $\mathcal{T}\ket{\psi_1(-t_1)}$,  $\mathcal{T}\ket{\psi_0(t=0)}$.
	\end{enumerate}
	
\end{enumerate}

\noindent It is important to note that the state $\ket{\psi_0(t=0)}$ and the time-reversed state $\mathcal{T}\ket{\psi_0(t=0)}$ will not, in general, be the same state. Momentum eigenstates (with $p \neq 0$) provide a simple example: the time-reverse of the state $\ket{p}$ is $\mathcal{T}\ket{p}=\ket{-p}$ (up to a phase), so the state and its time-reverse are orthogonal: $\expval{\mathcal{T}}{p}= 0$. Eigenstates of angular momentum observables for systems with non-integer spin exhibit the same behavior -- both the spin observables $\sigma_x, \, \sigma_y, \, \sigma_z$ and orbital angular momentum observables $L_x, \, L_y, \, L_z$.\footnote{This is a special case of the fact that the state of any degree of freedom that is \textit{odd} under time-reversal -- i.e. $\mathcal{T}^2\ket{\psi} = -\ket{\psi}$ -- is orthogonal to its time-reversed state \citep[section 3.2]{1987_sachs}.} One can specify conditions that ensure that a quantum state and its time-reverse are the same state (for example \citep[section 4]{2002_earman} or \citep[theorem 4.12]{2011_sakurai}), but such cases are the exception, not the rule.

Time-reversal invariance of the laws of microphysics is often identified as an obstacle to interpreting microphysics causally, but I think it is a red herring.\footnote{For two arguments for the same conclusion in classical and quantum statistical mechanics that have points of contact with the argument I offer here, but which are ultimately distinct, see \citep[chapter 4]{2007_maudlin} or \citep{2020_myrvold}.} Time-reversal invariance secures the following: if a sequence of states $\ket{\psi_a}, \, \ket{\psi_b}, \, \ldots, \, \ket{\psi_z}$ is dynamically allowed by unitary evolution under a Hamiltonian $H$, then a sequence of generally distinct states $\mathcal{T}\ket{\psi_z}$, $\ldots$,  $\mathcal{T}\ket{\psi_b}$,  $\mathcal{T}\ket{\psi_a}$ is also allowed by unitary evolution governed by the time-reversed Hamiltonian $\mathcal{T}H\mathcal{T}^{-1}=H$. 

For this to be an obstacle to interpreting microphysics causally it would have to be the case that given two states $\ket{\psi_a}$ and $\ket{\psi_z}$ related by some unitary evolution U(t), the time-reversal invariance makes it impossible to determine whether $\ket{\psi_a}$ is the cause of $\ket{\psi_z}$ or vice-versa. Time-reversal invariance is unecessary for this; as I discussed above, this obstacle arises from the invertibility of the time-evolution operators U(t) and that follows from Stone's theorem, whether the theory is time-reversal invariant or not. Time-reversal invariance is also in general insufficient for this, except in the trivial sense that if quantum theory is time-reversal invariant, that entails that one was able to define a Hamiltonian for the system and that, in turn, gets us back to the real obstacle to determining whether $\ket{\psi_a}$ is the cause of $\ket{\psi_z}$ or vice-versa: the invertibility of U(t) ensured by Stone's theorem.\footnote{For a related point about the relationship between time-reversal invariance and determinism to the past and future and additional discussion, see \citep[section 4]{2002_earman}.} If, on whatever basis, I were to claim to have good reason to believe that $\ket{\psi_a}$ is the cause of $\ket{\psi_z}$, it would do nothing to undermine my belief to tell me that the theory also allows for a physically distinct state $\mathcal{T}\ket{\psi_z}$ to be the cause of another physically distinct state $\mathcal{T}\ket{\psi_a}$. 

This is especially true for the situation I am considering in this paper: I am imagining that one \textit{knows} the two states $\ket{\psi_a}$ and $\ket{\psi_z}$. It is difficult to understand the belief that the difficulty with identifying whether $\ket{\psi_a}$ causes $\ket{\psi_z}$ or vice-versa stems from a fact about dynamically allowed sequences of states that are, in general, physically distinct and observationally distinguishable from $\ket{\psi_a}$ and $\ket{\psi_z}$. The real difficulty with identifying cause and effect in quantum theories stems from the invertibility of the dynamics: one cannot distinguish between the two possibilities $\ket{\psi_a} = \text{U(t)}\ket{\psi_z}$ and $\ket{\psi_z} = \text{U(t)}^{-1}\ket{\psi_a}$ in the absence of information about the temporal ordering of $\ket{\psi_a}$ and $\ket{\psi_z}$.

\section{Interactions Generate Entanglement}
	
Use of the Schmidt decomposition of quantum states is ubiquitous and useful when discussing entanglement. I will rely on it throughout the paper so what follows is a brief reminder about some of its relevant properties.\footnote{See \citep[section 2.5]{2010_nielsen_chuang} for a pedagogical presentation.}

Any pure state $\ket{\Psi}$ of a bipartite quantum system can be written as its Schmidt decomposition

\[
\ket{\Psi} = \sum_i \lambda_i \ket{a_i}_A \otimes \ket{b_i}_B
\]

\noindent where $\{\ket{a_i}_A\}$ and $\{\ket{b_i}_B\}$ form orthonormal bases for the Hilbert spaces $\mathcal{H}_A$ and $\mathcal{H}_B$, respectively. The coefficients $\lambda_i$ are called Schmidt coefficients and are non-negative real numbers that satisfy $\sum \lambda_i^2 = 1$. The \textit{Schmidt rank} of $\ket{\Psi}$ is the number of non-zero Schmidt coefficients $\lambda_i$; a state is entangled iff it has a Schmidt rank greater than 1 and it is ``fully'' entangled -- in a sense I will elaborate on in a moment -- if its Schmidt rank is equal to $\min\{\dim(\mathcal{H}_A), \, \dim(\mathcal{H}_B)\}$. For simplicity I will consider systems where $\dim(\mathcal{H}_A) = \dim(\mathcal{H}_B)$, but nothing in the paper depends on this.

If the Schmidt coefficients are non-degenerate, then the basis vectors $\ket{a_i}_A$ and $\ket{b_i}_B$ are unique (up to a phase) 
%compensating global phase factors (up to replacing $\ket{a_i}_A \otimes \ket{b_i}_B$ with  $e^{iphi}\ket{a_i}_A \$ e^{-i\phi}\ket{b_i}_B}$, for example) and 
so the Schmidt decomposition of a state $\ket{\Psi}$ is itself unique (up to a phase). If any of the Schmidt coefficients are degenerate then there is more freedom in the choice of basis vectors, but both the Schmidt rank and the values of the Schmidt coefficients are the same for any allowed choice of basis. A familiar example of this latter case is the EPR state. It can be Schmidt decomposed as 

\[
\ket{\text{EPR}} = \frac{1}{\sqrt{2}}\ket{0}_A\ket{0}_B + \frac{1}{\sqrt{2}}\ket{1}_A\ket{1}_B
\]

\noindent where the Schmidt coefficients are $\lambda_1 = \lambda_2 = \frac{1}{\sqrt{2}}$. One could equally well have written the state in the Schmidt decomposition 

\[\ket{\text{EPR}} = \frac{1}{\sqrt{2}}\ket{+}_A\ket{+}_B + \frac{1}{\sqrt{2}}\ket{-}_A\ket{-}_B
\]

\noindent where $\ket{\pm} = \frac{1}{\sqrt{2}}\ket{0} \pm \frac{1}{\sqrt{2}}\ket{1}$. The values of the Schmidt coefficients are unchanged, while the non-uniqueness of the basis reflects the degeneracy of those coefficients. 

This highlights a useful feature of the Schmidt decomposition: the Schmidt coefficients are invariant under unitary operations $U_A \otimes U_B$ that act on Alice and Bob's subsystems alone and, as an obvious corollary, the Schmidt rank is invariant too. This reflects the physical fact that entanglement between separated systems cannot be created by unitary operations performed locally on each system.\footnote{It also reflects the mathematical fact that the amount of entanglement between two systems is independent of a unitary change of basis.} As a result, the set of Schmidt coefficients $\lambda_i$ completely characterizes the entanglement of a bipartite system in a pure state and is sometimes referred to as the \textit{entanglement spectrum}.%following \citep{2008_li_haldane}. %In particular, the Schmidt coefficients of a bipartite pure state fully determine the value of a measure of entanglement used more frequently than the Schmidt rank, the entropy of entanglement.

This brings me to the sense in which a bipartite pure state with full Schmidt rank can be considered ``fully'' entangled. Consider an entangled bipartite system $\ket{\Psi} \in \mathcal{H}_{AB}$ with $\dim(\mathcal{H_A}) = \dim(\mathcal{H_B})=N$. If $\ket{\Psi}$ has Schmidt rank 1 then it is separable and Alice and Bob's measurement results will be uncorrelated for any measurements they perform. If $\ket{\Psi}$ has Schmidt rank \textit{N}, however, Alice and Bob's results will be correlated for \textit{every} measurement they perform in the same Schmidt basis on $\ket{\Psi}$: if Alice finds her system in state $\ket{a_k}$ then Bob will find his in state $\ket{b_k}$, and so on. This is the sense in which the two subystems are as entangled as they could be. %A more qualitiative way to put this is that since a state is separable iff it has Schmidt rank 1, a state with Schmidt rank N is as far from separable as allowed by the dimensionality of the Hilbert spaces $\mathcal{H}_A$ and $\mathcal{H}_B$.

The Schmidt rank may seem like a rather coarse entanglement measure. It does not distinguish states that an apparently more fine-grained measure of entanglement, like the commonly adopted entanglement entropy, would distinguish. Consider the density operators $\rho_1$ and $\rho_2$ corresponding to the entangled states

\[
\ket{\Psi_1} = \frac{1}{\sqrt{2}}\ket{11} + \frac{1}{\sqrt{2}}\ket{00}
\]

\noindent and 

\[
\ket{\Psi_2} = \sqrt{1 - \varepsilon}\ket{11} + \sqrt{\varepsilon}\ket{00}
\]

\noindent defined on $\mathcal{H}_{AB}$ with $\dim(\mathcal{H}_A) = \dim(\mathcal{H}_B) = 2$. Both $\rho_1$ and $\rho_2$ have the full Schmidt rank 2. However, the entanglement entropy of a density operator 

\[
EE(\sigma) = -\Tr(\sigma \log\sigma) = -\sum_i \lambda^2_i \log\lambda_i^2
\]

\noindent distinguishes the two: $EE(\rho_1) =1$ while $EE(\rho_2)$ is $\order{\varepsilon}$.\footnote{The squares of the Schmidt coefficients $\lambda_i^2$ in the Schmidt decomposition of a bipartite pure state $\ket{\Psi}$ are the eigenvalues of the density operator $\sigma = \dyad{\Psi}$. A convenient fact about the Schmidt decomposition is that the $\lambda_i^2$ are \textit{also} the eigenvalues of each of the reduced density operators $\rho_A$ and $\rho_B$ representing each entangled subystem. Since these eigenvalues fully determine the entanglement entropy, the Schmidt decomposition reveals that one can calculate the entanglement entropy of the bipartite system in two equivalent ways: by computing the Shannon entropy of the probability distribution generated by the square of the coefficients of $\sigma$ for the full bipartite system or, as is more common, computing the von Neumann entropy of the reduced density operators $\rho_A$ or $\rho_B$.} Why not use the apparently more fine-grained measure of entanglement?

My reasons are partially pragmatic. The first such reason is that such a measure is more fine-grained than I need for present purposes since the topological, measure-theoretic, and dynamical facts about quantum states that I invoke in the remainder of this section can be proved using only information about the Schmidt rank. The second is that there is virtue in adopting a measure of entanglement can be extended to multipartite systems and to mixed states. As a measure of entanglement, the Schmidt rank generalizes naturally to such cases: the Schmidt measure \citep{2001_eisert_briegel}, \citep[section II.c]{2004_hein_eisert_briegel} or the Schmidt number \citep{2000_terhal_horodecki,2001_bruss,2011_sperling_vogel} are generalizations of the Schmidt rank to multipartite pure and mixed states. In fact, the availability of the Schmidt measure will be important in section 5 when I consider a multipartite system of \textit{N} qubits. A third reason, closely related to the second, is that the Schmidt measure has been used to give a definition of the algorithmic information content (also called the algorithmic complexity or the Kolmogorov complexity) of quantum states \citep{2005_mora_briegel,2006_mora_briegel}. I will make use of this connection when I turn to causal inference in section 5.

My reasons are not entirely pragmatic, however, and there are two conceptual issues worth highlighting before moving on. There is a sense in which the Schmidt rank and its generalizations are not measures of precisely the same property as the entanglement entropy. Consider the following remarks from \citep{1998_preskill} and \citep{2001_bruss}:

\begin{quote}
	So a number used to quantify entanglement ought to have the property that local operations do not increase it. An obvious candidate is the Schmidt [rank], but on reflection it does not seem very satisfactory. Consider
	
	\[
	\ket{\Psi_\varepsilon} = \sqrt{1-2\left|\varepsilon^2\right|}\ket{00} + \varepsilon\ket{11} + \varepsilon\ket{22}
	\]
	
	which has Schmidt [rank] 3 for any $\left|\varepsilon\right| > 0$. Should we really say that $\ket{\Psi_\varepsilon}$ is ``more entangled'' than $\ket{\phi^+} = \frac{1}{\sqrt{2}}(\ket{11} + \ket{00})$? Entanglement, after all, can be regarded as a resource -- we might plan to use it for teleportation, for example. It seems clear that $\ket{\Psi_\varepsilon}$ (for $\left|\varepsilon\right| \ll 1$) is a less valuable resource than $\ket{\phi^+}$ \citep[chapter 5.5]{1998_preskill}.\footnote{The Schmidt \textit{rank} of a bipartite pure state is sometimes referred to interchangeably as its Schmidt \textit{number}. This is unfortunate since I will occasionally mention a quantity introduced by \citep{2000_terhal_horodecki} that they call the Schmidt number, which is an extension of the Schmidt rank to mixed states. I've altered Preskill's terminology to cohere with mine in this paper.}
\end{quote}

\noindent Bru{\ss} draws a similar distinction as Preskill while introducing a method for determining the Schmidt number of an arbitrary mixed state:

\begin{quote}
	A slightly different question from ``how much entangled is a state $\rho$?'' can be addressed via the generalization of entanglement witnesses to so-called Schmidt witnesses. They give an answer to the question ``how many degrees of freedom are entangled in $\rho$?'' This corresponds to a \textit{finer classification} of entangled states \citep[section V.C]{2001_bruss}.
\end{quote}

\noindent It is not hard to find similar remarks in the literature on entanglement which suggest that rather than the entanglement entropy being a finer measure than the Schmidt rank or vice versa, the two instead capture subtly different aspects of entanglement.\footnote{For example, see the remarks in \citep{2001_sanpera_bruss_lewenstein} or, more substantively, the proof in \citep{2013_den_nest} mentioned below. This is not to say that there is never any relationship between the entanglement entropy and the Schmidt rank; for example, see \citep[section III]{2011_sperling_vogel}. For philosophical discussion of some of the multiple notions of entanglement see \citep{2015_earman}.} Heuristically, the entanglement entropy quantifies how \textit{strongly} subsystems are entangled while the Schmidt rank captures how \textit{broadly} the subsystems are entangled. This distinction can track important practical differences in certain contexts: for example, it was proved in \citep{2013_den_nest} that the Schmidt rank is an informative measure of entanglement for determining the potential speedup of a quantum computation over a classical one, but any entanglement measure that is continuous and vanishes on product states -- like the entanglement entropy -- is not.%It is at minimum unclear that the assumption that instigated this brief defense of the Schmidt rank as a measure of entanglement -- that the entanglement entropy is simply a more fine-grained measure of entanglement than the Schmidt rank -- is even accurate.

The second conceptual issue concerns a connection between measures of entanglement and measures of information that I will exploit when discussing causal inference.\footnote{See \citep[chapters 2.2 \& 3.6]{2013_timpson} for a clear presentation and philosophical discussion of the interpretation of the Shannon and von Neumann entropies as measures of information.} The von Neumann entropy -- what I have thus far been calling the entanglement entropy -- can also be given an information-theoretic interpretation as a quantum analogue of Shannon information. The Shannon information does not quantify the information content of an individual message, but rather characterizes an \textit{ensemble} of messages produced by a specified source that produces each bit $x_i$ with some probability $p_i$. The Shannon information quantifies the minimum number of bits required to encode an arbitrary message selected from an ensemble of typical messages produced by the source. This is also true of the von Neumann entropy; a quantum source prepares qubits in pure states $\ket{x_i}$ with probabilities $p_i$ and the entanglement entropy quantifies the minimum number of qubits required to transmit a typical message selected from the ensemble of messages produced by the specified quantum source. %This manifests in the fact that both the Shannon entropy of a random variable X and the von Neumann entropy of a density operator $\sigma$ are functions of the probability distribution over the possible values of X, or pure quantum states, determined by a specified classical (or quantum) source.

The causal inference methods I will use in the subsequent sections rely on the notion of \textit{algorithmic} information.\footnote{See \citep{2010_grunwald_vitanyi} for a helpful comparison of Shannon information and algorithmic information.} This is not a measure of the amount of information required to describe a typical object in a specified ensemble but of the ``intrinsic'' amount of information contained in the object considered on its own. I will have more to say about algorithmic information in section 4 but what matters presently is that the algorithmic information content of an object is \textit{not} a statistical property of an ensemble containing that object. This suggests that the entanglement entropy cannot provide a satisfactory measure of the algorithmic information content of an individual pure quantum state. The Schmidt rank of a quantum state is a property of that state alone and is thus a more appropriate measure for drawing a connection between entanglement and the algorithmic information content of a quantum state. (They do not motivate their choice in this way but this does make the use of the Schmidt measure by \citep{2005_mora_briegel,2006_mora_briegel} a conceptually natural choice for explicitly connecting entanglement and the algorithmic information content of a quantum state.)

With those justifications provided, I will now make use of the Schmidt rank to introduce some topological, measure-theoretic, and dynamical facts about the ubiquity of entangled pure states in the state space $\mathcal{H}_{AB}$ of a bipartite quantum system. All of these facts are well-known; my aim in rehearsing them is to lay out the physical grounds of the causal asymmetry for which I argue at the end of this section, as well as to make maximally plausible the similar conclusions I draw using causal inference methods in section 5. If this discussion results in those conclusions seeming inevitable then it will have been successful.

First consider the case where  $\mathcal{H}_{AB}$ is finite-dimensional. Recall that the rank of a matrix is the number of linearly independent rows (or columns) it contains and that a matrix \textit{M} of less-than-full rank has $\det(M)=0$. This means that the density matrix corresponding to any bipartite pure state of less-than-full Schmidt rank has determinant equal to zero. One can use this fact to show that the set of pure states in $\mathcal{H}_{AB}$ with less-than-full Schmidt rank is nowhere-dense in the set of all pure states in the topology induced by any norm on the space.\footnote{See \citep{2005_brock} for an elementary proof. The fact that all norms on a finite-dimensional vector space are equivalent justifies the statement that this is true for ``any norm'' on $\mathcal{H}_{AB}$.} This means that each pure state in $\mathcal{H}_{AB}$ of less-than-full Schmidt rank is enveloped by a ball of fully entangled pure states; equivalently, each pure state of less-than-full Schmidt rank is a limit point of a sequence of fully entangled pure states. 

The set of pure states of less-than-full Schmidt rank is also sparse measure-theoretically: it has measure zero in the set of all pure states in $\mathcal{H}_{AB}$.\footnote{In the finite-dimensional case, the restriction to pure states is important: for \textit{mixed} states of a bipartite system, the separable states are no longer as sparse, and the set of separable mixed states always contains an open ball around the maximally mixed state \citep{1998_zyczkowski_horodecki_sanpera_lewenstein}. For an infinite-dimensional Hilbert space, this is no longer true: as \citep{1999_clifton_halvorson} showed, the set of mixed states is nowhere dense in the set of all states, in the topology induced by the trace-norm.} This is well-known but I couldn't find a proof to cite, so here is a proof sketch. Since $\mathcal{H}_{AB} \cong \mathbb{C}^{\text{N} \times \text{N}}$, the determinant $\det \colon \mathbb{C}^{\text{N}\times \text{N}} \to \mathbb{C}$ is a polynomial in the complex entries of an $\text{N}\times \text{N}$ matrix and any complex polynomial is a holomorphic function. The zero set \textit{Z(f)} of any holomorphic function -- the set of points in the domain of \textit{f} on which that function equals zero -- is at most countable \citep[theorem 10.18]{1987_rudin} and so the determinant equals zero on at most countably many points in $\mathbb{C}^{\text{N} \times \text{N}}$. The determinant of a density matrix vanishes if and only if it has less-than-full Schmidt rank so there can be at most countably many such density matrices. Using $\mathbb{C}^{\text{N} \times \text{N}} \cong \mathbb{R}^{2\text{N}\times 2\text{N}}$ we can conclude that the set of density matrices with less-than-full Schmidt rank has Borel measure zero.

In light of the sparseness of separable pure states it may seem obvious that non-trivial time evolution on $\mathcal{H}_{AB}$ -- any time evolution that includes interactions between Alice and Bob's subsystems -- has to take any pure state $\ket{\psi}_{AB}$ that is not fully entangled into a fully entangled state, and that the curve traced out by U(t)$\ket{\psi}_{AB}$ will consist almost entirely of fully entangled states.\footnote{I will drop the ``non-trivial'' qualifier for the remainder of the paper. Considering time evolutions that are non-trivial in this sense is equivalent to requiring that the time evolution operator U(t) does not factorize into a product  $\text{U(t)}_A \otimes \text{U(t)}_B$ of operators evolving Alice and Bob's subsystems independently. Any unitary operator that factorizes in this way cannot change the Schmidt coefficients of a quantum state, as I mentioned previously, and so cannot create entanglement between subsystems.} This is true.\footnote{A general argument is sketched in \citep[section 6.1]{2013_binney_skinner}. There are an immense number of concrete examples of dynamically generated entanglement; see, for instance, the discussions in many-body physics \citep{2006_calabrese_cardy,2006_eisert_osborne,2008_amico_et_al}, quantum field theory \citep{2016_peschanski_seki,2017_cervera_lieta_et_al,2017_kharzeev_levin}, and non-relativistic quantum mechanics \citep{2004_mishima_hayashi_lin,2017_schroeder}.} In fact, the previous discussion ensures that any initial bipartite state with less-than-full Schmidt rank will, after an infinitesimal time interval, develop into a state with full Schmidt rank and will then remain in a state of full Schmidt rank for all subsequent times except for a set of measure zero. 

This should not be surprising at this point. The topological sparseness of pure states of less-than-full Schmidt rank means that there is nowhere else for those states to evolve except into states with full Schmidt rank, at least for infinitesimal time evolution, and their measure-theoretic sparseness means that any curve through $\mathcal{H}_{AB}$ generated by time evolution for \textit{any finite time interval} will pass through a state of less-than-full Schmidt rank in at most countably many instants. The upshot of this is that if one chooses an arbitrary state $\ket{k}$ of less-than-full Schmidt rank \textit{k} and evolves it for any finite time \textit{t}, except for set of isntants of measure zero the state U(t)$\ket{k}$ will be a state of full Schmidt rank.

Now suppose instead that one is interested in the time evolution of an arbitrary state $\ket{N}$ of full Schmidt rank \textit{N}. Are there any time evolution operators U(t) such that U(t)$\ket{N}$ has less-than-full Schmidt rank? Since specifying a Hilbert space representation for a quantum system requires fixing a Hamiltonian $H$, the only freedom in choosing U(t) comes from fixing \textit{t}.\footnote{This is clearest in the algebraic approach to quantum theories, where a quantum system is defined by a C$^*$ algebra. The GNS reconstruction theorem enables one to move from the abstract C$^*$ algebra to a Hilbert space representation for the system, but this requires data about the full C$^*$ algebra, including the Hamiltonian.}. Thus this question is equivalent to asking how many instants $t_i$ there are such that U($t_i$)$\ket{N}$ has less-than-full Schmidt rank.

One such operator can obviously be reverse-engineered. Find a state of less-than-full Schmidt rank $\ket{k}$ such that U(t)$\ket{k} = \ket{N}$. Applying $\text{U(t)}^{-1}$ to $\ket{N}$ will then produce $\ket{k}$, a state of less-than-full Schmidt rank. Are there more such operators? 

%Let $\tau_R$ be the recurrence time for the system.\footnote{See \citep{2015_wallace} for proofs that in any finite-dimensional quantum system, and any infinite-dimensional quantum system that satisfies modest constraints on the Hamiltonian, there exists a single time $\tau_R$ such that after $\tau_R$ \textit{every} state in $\mathcal{H}_{AB}$ will have recurred.} The operator U($\tau_R$ - t) will do the trick: $\text{U}(\tau_R - \text{t}) \ket{N} = \text{U}(\tau_R)[\text{U(-t)} \ket{N}] = \text{U}(\tau_R)\ket{\phi} = \ket{\phi}$. 

There can be at most countably many. Let $\tau_R$ be the recurrence time for the system and consider the operator U($\tau_R$).\footnote{See \citep{2015_wallace} for proofs that in any finite-dimensional quantum system, and any infinite-dimensional quantum system that satisfies modest constraints on the Hamiltonian, there exists a single time $\tau_R$ such that after $\tau_R$ \textit{every} state in $\mathcal{H}_{AB}$ will have returned to a state arbitrarily close to itself.} Suppose that one starts with the state $\ket{N}$ and that the dynamics are such that $\ket{N}$ visits \textit{every} state of less-than-full Schmidt rank in $\mathcal{H}_{AB}$ between $t = 0$ and $t = \tau_R$. Then the curve $\text{U}(\tau_R)\ket{N}$ will visit countably many states of less-than-full Schmidt rank on its tour through $\mathcal{H}_{AB}$. Any finite time interval \textit{t} can be split up into at most \textit{n} intervals of length $\tau_R$ (plus some remainder \textit{q}) so we can write $\text{U(t)} = \text{U}(n\tau_R + q)$ and, since a finite union of countable sets is itself countable, the state $\ket{N}$ evolves into a state of less-than-full Schmidt rank for at most countably many instants. This means that the set of times $t_i$ for which U($t_i$)$\ket{N}$ has less-than-full Schmidt rank has measure zero. 

%Give concrete problem on qubits from StackExchange/Nielsen \& Chuang

Recall that I began with a simple problem: if one is given two bipartite pure states $\ket{\alpha}$ and $\ket{\beta}$ and knows that they are related by unitary time evolution, can they determine whether $\ket{\alpha}$ was the cause of $\ket{\beta}$ or vice versa? My discussion of the sparseness of pure states of less-than-full Schmidt rank in $\mathcal{H}_{AB}$ lays the groundwork for an answer. Suppose that either of the following is true:\footnote{\label{NMSR}Note that this excludes the case where both $\ket{\alpha}$ and $\ket{\beta}$ are entangled but neither has full Schmidt rank. I'm not aware of any demonstration that within the set of bipartite pure states of less-than-full Schmidt rank, the states of lower rank are less prevalent than those of higher rank. The causal inference methods used in section 5 will do better for this case.}

\begin{enumerate}
\item \textit{The state $\ket{\alpha}$ is separable (has Schmidt rank 1) and $\ket{\beta}$ is entangled (has Schmidt rank $>$1)}; \begin{center}or\end{center}
\item \textit{Both of the states $\ket{\alpha}$ and $\ket{\beta}$ are entangled but $\ket{\alpha}$  has less-than-full Schmidt rank and $\ket{\beta}$ has full Schmidt rank.}
\end{enumerate}

\noindent It is then overwhelmingly likely that $\ket{\beta}$ was caused by time-evolving $\ket{\alpha}$ rather than vice versa. 

The intuitive reason for this is that it is overwhelmingly likely that a state of low Schmidt rank will evolve into a state of higher Schmidt rank but the converse is extremely unlikely. If the state $\ket{\alpha}$ is the cause then it is almost certain that the state $\text{U(t)}\ket{\alpha}$ would have higher Schmidt rank, for an arbitrarily chosen \textit{t}. One does not have to know anything about the dynamics -- about how $\ket{\alpha}$ would evolve over different intervals \textit{t} -- to be essentially certain that the resultant state will have full Schmidt rank. If one evolved the state $\ket{\alpha}$ for $t + dt$ or $t + 10^{23}$, or $t + \pi t$, or $\ldots$ the resultant state would not be $\ket{\beta}$, but one can be essentially certain that it would be a fully entangled state. That it happens to be $\ket{\beta}$ in particular reflects the unmysterious fact that one evolved the state $\ket{\alpha}$ for \textit{t} rather than some other interval of time. In short, identifying $\ket{\alpha}$ as the cause renders the effect $\ket{\beta}$ utterly unsurprising.

If instead one adopts $\ket{\beta}$ as the cause, then the effect $\ket{\alpha}$ is rendered \textit{fantastically} surprising. It is impossibly unlikely that $\ket{\beta}$ just happened to evolve for precisely an interval \textit{t} lying in the measure-zero set of times that would result in $\text{U(t)}\ket{\beta}$ having less-than-full Schmidt rank. It suggests an incredible degree of fine-tuning; indeed, it is nearly impossible to imagine a human experimenter who was \textit{trying} to accomplish such a precise fine-tuning doing so successfully. Even with complete knowledge of both the state $\ket{\beta}$ and its behavior under the dynamics -- how it would evolve under U(t) for different \textit{t} -- and the explicit goal of producing a state of lower-Schmidt rank, they would have to control the experiment with a level of precision typically reserved for deities.\footnote{\label{Schwinger}An observation about the impracticality of arranging a similar evolution was made in a different context by \citep{1988_schwinger_englert_scully}. They consider sending a spin-1/2 particle in a $\sigma_x$ eigenstate through a Stern-Gerlach device set to measure $\sigma_z$, which will split the incoming beam into a superposition of the two eigenstates of $\sigma_z$. They investigate the degree of precision with which an experimenter would need to control the magnetic field in the Stern-Gerlach device to ensure that the spin-1/2 particle returns to its original $\sigma_x$ eigenstate after passing through the Stern-Gerlach device. %This is equivalent to evolving the $\sigma_x$ eigenstate from $t=0$ to $t=T$ under the Hamiltonian describing its interaction with magnetic field in the Stern-Gerlach device, applying a time-reversal operation, evolving the state from $t=T$ to $t=2T$, and applying a time-reversal operation again. 
They show that an \textit{exact} return to the original $\sigma_x$ eigenstate is unattainable in practice and that even to reproduce the original state with 99\% accuracy
% (with accuracy measured by deviation from $\left |1 - \expval{\sigma_x}(2T)\right|$) 
would require the ability to control the gradient of the macroscopic magnetic field in the Stern-Gerlach device to \textit{at least} 5 decimal places.} In fact, insofar as one thinks that a cause ought to explain its effects then identifying $\ket{\beta}$ as the cause is unacceptable: it not only fails to explain $\ket{\alpha}$ in any meaningful sense but also raises additional explanatory concerns more urgent than those with which one began.

Some readers may have noticed a parallel with classical statistical mechanics: precisely this form of reasoning has been used in that context to argue that there is an identifiable explanatory asymmetry between two microstates despite the time-symmetry of the dynamical laws describing their time evolution. For example, \citep[p. 132]{2007_maudlin} introduces the problem as follows:

\begin{quote}
$\ldots$ postulate a macroscopically atypical [low entropy] but microscopically typical [chosen at random from microstates compatible with the atypical macrostate] state, plus the laws, and one can explain the macroscopically typical [high entropy] but microscopically atypical [evolves to a lower entropy macrostate in one temporal direction] state from them: the latter was generated from the former by means of the operation of the laws. But equally: postulate a macroscopically typical but microscopically atypical state at one end, plus the laws, and one can `generate' a macroscopically atypical but microscopically typical state from them. Pick one end, add the laws, and you can explain the other end: which end you pick as \textit{explanans} and which as \textit{explanandum} is up to you.
\end{quote}

\noindent Maudlin argues that this apparent explanatory symmetry is illusory, invoking a principle that lies at the foundation of the causal inference methods I discuss in section 4. He points out that the asymmetry is hiding in the way that one specifies the \textit{explanans} microstate and the \textit{explanandum} microstate in the two candidate explanations. Consider explaining an atypical microstate in a high-entropy macrostate by demonstrating that it resulted from the time evolution of a typical microstate in a low-entropy macrostate. The meaning of ``typical'' used to describe the \textit{explanans} is measure-theoretic: the low-entropy macrostate is atypical because it occupies a small volume of the system's phase space and the microstate is typical because it was chosen at random from that low-entropy macrostate. Importantly, one can entirely specify an \textit{explanans} that will produce the desired \textit{explanandum} while knowing basically nothing about the dynamics governing the system.%\footnote{I say \textit{basically} nothing to allow for information like certain generic mathematical conditions, such as continuity or smoothness properties, that a function may have to satisfy to be a well-defined input into the dynamics or to have a unique solution.}

This is no longer true if one tries to invert the order of explanation. If one chooses as the \textit{explanans} an atypical microstate in a high-entropy macrostate, the meaning of ``atypical'' becomes crucially different. ``Atypical'' now means \textit{dynamically} atypical: the microstate is atypical in the new sense if its time evolution takes it into a low-entropy macrostate. Such atypical microstates are scattered throughout the volume of the high-entropy macrostate and share no common property that could be used to identify them except their dynamical behavior. Specifying an atypical microstate in this sense requires thorough knowledge of the dynamics governing its evolution. Not only can such microstates only be specified by describing their dynamical behavior, but their identity \textit{as} atypical microstates depends very sensitively on the specific form of the dynamics governing the system. As Maudlin points out, 

\begin{quote}
$\ldots$ a slight modification of the dynamical laws would lead to essentially no change in which initial states are macroscopically atypical, in that they have low entropy, \textit{but would completely alter the set of atypical high-entropy states whose time evolution in either direction leads to low entropy} \citep[p. 133]{2007_maudlin} [my emphasis].
\end{quote}

\noindent The explanatory asymmetry comes from this asymmetry in how we have to specify the \textit{explanans} and the \textit{explanandum} in the two cases. In one direction, one can randomly pick a microstate from a low-entropy macrostate, knowing essentially nothing about the specific dynamics governing the system, and then show that those dynamics will evolve that \textit{explanans} microstate into a microstate lying in a high-entropy macrostate. In the other direction, one needs detailed knowledge about the specific form of the dynamics to even know what counts as an atypical microstate, let alone to select as their \textit{explanans} an atypical microstate that the dynamics will evolve into a low-entropy macrostate. Maudlin argues that this is no explanation at all: it amounts to explaining the system's evolution into a low-entropy macrostate by saying that the initial conditions were such that they would evolve into a low-entropy macrostate, given the dynamics governing the system. One might as well ascribe such microstates a \textit{virtus entropia}.\footnote{Sklar expresses a similar dissatisfaction with such candidate explanations of why subsystems of a larger system obey the second law: ``$\ldots$ we would simply posit an initial state that gives rise to parallel entropic increase in branch systems with each other and with the main system. But to characterise the state in that way would, of course, not be offering us an explanation of the sort we expected. It would be one thing to be able to characterize the initial state in some simple way$\ldots$ and be able to derive the Second Law from that. But to derive the Second Law from a bald assertion that ``initial conditions were such that they would lead to Second Law behavior'' hardly seems of much interest'' \citep[p. 330]{1993_sklar}.}

The parallel strategy for using entanglement to identify explanatory asymmetries between quantum states is clear. The set of all states in $\mathcal{H}_{AB}$ can be partitioned into disjoint subsets containing states with Schmidt rank 1, 2, $\ldots$, $\dim(\mathcal{H}_{AB})$. We know from the above discussion that almost all states in $\mathcal{H}_{AB}$ will be in the subset of states with full Schmidt rank; call a quantum state in $\mathcal{H}_{AB}$ ``macroscopically atypical'' if it lies in any set of states with less-than-full Schmidt rank. A state \textit{within} that set is ``microscopically typical'' if it is chosen from that set at random. A state is ``macroscopically typical'' in $\mathcal{H}_{AB}$ if it is in the set of states with full Schmidt rank; a state within that set is ``microscopically atypical'', given a specific time-evolution operator U($t_i$), if it evolves into a state of less-than-full Schmidt rank. 

The discussion above establishes that time evolution from full Schmidt rank to less-than-full Schmidt rank is highly non-generic; the only way to specify an \textit{explanans} state $\ket{\phi}$ that would do so under U($t_i$) requires extremely detailed knowledge of both the details of the Hamiltonian $H$ appearing in U($t_i$) and the duration of time evolution $t_i$. Just like in the classical statistical mechanical case, ``explaining'' a state $\ket{\phi_1}$ of less-than-full Schmidt rank  as resulting from the time evolution of a state $\ket{N}$ of full Schmidt rank amounts to saying that $\ket{\phi_1}$ occurred because the initial condition $\ket{N}$ was such that it would evolve to a state of lower Schmidt rank, given the dynamics U($t_i$).

Maudlin extracts from his discussion a principle for evaluating explanations, though he doesn't dwell on it:\footnote{Although Maudlin doesn't dwell on it, \citep{2020_woodward} has given an extended and enlightening examination of the very similar principle that forms the foundation of the causal inference methods I describe in section 4.}

\begin{quote}
The problem is this: in order to
account for the universe as we see it, we need more than the laws: we need
a constraint on one of the boundaries. That constraint, together with the
operations of the laws, then suffices to account for the nature of the other
boundary. But in order for this to work [be explanatory] \textit{the constraint must itself be specifiable independently of what will result from the operation of the laws} \citep[p. 132]{2007_maudlin}.
% emphasis in original
\end{quote}

\noindent I think there is an argument to be made at this point that as long as two quantum states satisfy one of the two conditions outlined above, my initial goal of identifying a causal asymmetry between quantum states related by time-symmetric dynamics has been accomplished. Recall that I have adopted a minimal interventionist notion of causation in this paper: X is a cause of Y iff there is an intervention that can be performed on the value of X, while holding all other variables fixed, that produces a change in the value of Y. Two quantum states $\ket{\alpha}$ and $\ket{\beta}$ related by a unitary time evolution U(t) satisfy this condition, but they satisfy it for both candidate causal orderings: intervening to set $\ket{\alpha}$ to $\ket{\alpha'}$ will produce a change in $\ket{\beta}$, since U(t)$\ket{\alpha'} \neq \ket{\beta}$. But this is also true if the candidate causal order is inverted: intervening to set $\ket{\beta}$ to $\ket{\beta'}$ will produce a change in $\ket{\alpha}$ since U(t)$^{-1}$$\ket{\beta'} \neq \ket{\alpha}$. 

Some additional information is required to break the symmetry. The above discussion suggests that, under conditions widely satisfied in quantum theories, one has such information: one should identify the ``cause'' state as the state that can produce the effect state \textit{and} can be specified without any reference to its behavior under the dynamics U(t). In the case under consideration, this amounts to identifying the ``cause'' state as either (1) the state with Schmidt rank 1, if the other state has Schmidt rank $>$ 1 or (2) the state with less-than-full Schmidt rank, if the other state has full Schmidt rank. Many people have claimed that the time-symmetry of the dynamics governing microscopic systems makes any such asymmetry impossible; I have argued that this is not correct.

That said, I want to emphasize some limitations of my discussion thus far. First, I have discussed only microscopic systems \textit{par excellence}: pure states of bipartite systems related by unitary evolution. Second, conditions (1) or (2) are satisfied for a pair of states only when one of those states comes from a set of measure zero: the set of states of less-than-full Schmidt rank. It is true that such states, particularly product states, play a more significant role in foundational and mathematical discussions of quantum mechanics than their sparseness in $\mathcal{H}_{AB}$ alone might suggest, but they are sparse nonetheless. I invoked an analogy with the case of classical statistical mechanics above, but this is an important disanalogy: the measure of the set of low-entropy microstates is exponentially small and vanishes as the size of the system $N\to 0$, but it is non-zero for any finite system; for estimates and discussion of classical and quantum multipartite systems (including exceptions to this estimate), see \citep{2015_goldstein_et_al} and \citep[section 7]{2017_goldstein_et_al}. I will return to this disanalogy in section 5.

The final limitation is that while conditions (1) and (2) show that one can, in principle, identify a causal asymmetry in quantum theories, they are of little practical value for the real-world epistemology of causation. The argument presented in this section does identify an asymmetry present in the \textit{mathematical foundations} of quantum theories, but it relies on a standard of precision that unsatisfiable in practice. I have been considering states that are \textit{exactly} product states or, more generally, states whose Schmidt decomposition has \textit{exactly} \textit{k} non-zero coefficients. This presumption of precision is often made without comment in foundational discussions but is never achievable -- or at least never verifiably achievable -- in any real world situation. At best, one can verify that a state is \textit{approximately} a product state or that a state has \textit{k} Schmidt coefficients larger than some $\varepsilon > 0$. However, the topological sparseness of the states of less-than-full Schmidt rank means that if $\ket{\psi}$ has less-than-full Schmidt rank, any state $\ket{\phi}$ that is within $\varepsilon$ of $\ket{\psi}$ in the Hilbert space norm will have full Schmidt rank. Conditions (1) and (2) no longer identify a causal asymmetry if modified to 

\begin{enumerate}
	\item[]$1_\varepsilon$: The state $\ket{\alpha}$ is \textit{within $\varepsilon$ of a state that is} separable (has Schmidt rank 1) and $\ket{\beta}$ is \textit{within $\varepsilon$ of a state that is} entangled (has Schmidt rank $>$1), or;
	\item[]$2_\varepsilon$: Both of the states $\ket{\alpha}$ and $\ket{\beta}$ are entangled but $\ket{\alpha}$  is \textit{within $\varepsilon$ of a state that has} less-than-full Schmidt rank and $\ket{\beta}$ is \textit{within $\varepsilon$ of a state that has} full Schmidt rank.
\end{enumerate}

In section 5 I will show that one can do better than this while still making use of a principle much like the one Maudlin employed to identify an explanatory asymmetry in classical statistical mechanics and which I have used in quantum theories. Such a principle provides part of the foundation of a set of mathematical and conceptual methods for causal inference that have proven empirically reliable in a number of disparate domains. My aim is to make use of those methods to give a formal demonstration that the presence of entanglement can be used to infer causal direction in quantum theories under conditions that are less restrictive than those invoked in this section.

\section{Causal Inference Methods}

I began with the problem of determining whether, and under what conditions, one could identify a causally relevant asymmetry between quantum states related by unitary evolution. Faced with such a problem, it would be natural to turn to the tools of causal inference. Unfortunately, if one takes up commonly used tools to infer causal relationships from observational data that rely solely on conditional statistical independence, such as those in \citep[chapter 2]{2009_pearl}, they will find those tools inadequate for the problem at hand. The reason for this is fairly straightfoward.

Any causal graph relating classical statistical variables V$_1$, V$_2$, $\ldots$, V$_\text{N}$ defines a joint probability distribution P(V$_1$, V$_2$, $\ldots$, V$_\text{N}$) over those variables.\footnote{\label{fnjointdist}The restriction to \textit{classical} statistical variables is important because a joint probability distribution for non-commuting variables, like conjugate observables in quantum mechanics, is generally not well-defined. If one restricts to quantum observables that commute then one can define a joint probability distribution over their possible values. See \citep{1982_fine} for a review and connection to hidden variable theories.} Consider a simple graph relating V$_1$, V$_2$, and V$_3$:

\begin{center}
	\begin{tikzpicture}
		% x node set with absolute coordinates
		\node[state] (z) at (0,0) {$V_3$};
		
		% y node set relative to x.
		% Locations can be:
		% right,left,above,below,
		% above left,below right, etc
		\node[state] (y) [below right =of z] {$V_2$};
		
		% y node set relative to x.
		% Locations can be:
		% right,left,above,below,
		% above left,below right, etc
		\node[state] (x) [below left=of z] {$V_1$};
		
		% Directed edge
		\path (x) edge (z);
		
		% Directed edge
		\path (y) edge (z);
		
		% Bidirected edge
		%\path[bidirected] (x) edge[bend left=60] (y);
	\end{tikzpicture}
\end{center}

\noindent This graph defines a set of possible joint probability distributions P(V$_1$, V$_2$, V$_3$). Suppose that all causal relations are deterministic; then possible joint distributions could be P(V$_1$=1, V$_2$=0, V$_3$=1)=1, P(V$_1$=0, V$_2$=0, V$_3$=1)=0, and so on. Causal inference problems begin with a joint probability distribution over variables V$_1$, V$_2$, $\ldots$, V$_\text{N}$ and attempt to reconstruct the causal graph representing the true causal structure that generates the given probability distribution. 

For each candidate causal graph, one typically requires that a probability distribution satisfies two conditions relative to that graph: the Causal Markov Condition  and a faithfulness condition.\footnote{See, for example, \citep[chapter 6.5]{2017_peters_janzing_scholkopf}, \citep[chapter 3]{2000_spirtes_glymour_scheines}, or \citep[chapter 2.4]{2009_pearl}, where the faithfulness condition is called ``stability''.}  The Causal Markov Condition is a constraint on how the joint probability distribution P(V$_1$, V$_2$, $\ldots$, V$_\text{N}$) factorizes into a product of conditional dependence relations according to a candidate causal graph. Speaking loosely, one requires that if all of the direct causes of a variable \textit{O} in the causal graph are specified, then one cannot learn any additional information about the value of \textit{O} from any variable that is not itself a descendent of  \textit{O}. Formally, one requires that any candidate causal graph satisfy: 

\[
P(\text{V}_1, \, \ldots , \, \text{V}_\text{N}) = \prod_{k=1}^{n} P(\text{V}_\text{k} \mid PA(\text{V}_\text{k}))
\]

\noindent where $PA(\text{V}_\text{k})$ represents the direct causes (or ``parents'') of the variable V$_k$. This captures the requirement that all of the conditional dependence relations in the joint distribution P(V$_1$, V$_2$, $\ldots$, V$_\text{N}$) be accounted for by causal relations between V$_1$, V$_2$, $\ldots$, V$_\text{N}$ in the causal graph.

Faithfulness requires that \textit{only} the conditional dependence relations in the joint probability distribution are reflected in the causal relations between V$_1$, V$_2$, $\ldots$, V$_\text{N}$ in the causal graph. The idea is easiest to illustrate with an example. Suppose one knows the joint distribution P(V$_1$, V$_2$, V$_3$) and suppose that V$_1 \indep \text{V}_2$, but V$_1 \slashed{\indep} \text{V}_2 \, \mid \text{V}_3$. (Notation: $\indep$ indicates statistical independence and $\slashed{\indep}$ indicates statistical dependence.) This is consistent with both of the following causal graphs: 

\vspace{\baselineskip}

\begin{center}
	
	\begin{minipage}{.25\paperwidth}

		\begin{tikzpicture}
			% x node set with absolute coordinates
			\node[state] (z) at (0,0) {$V_3$};
			
			% y node set relative to x.
			% Locations can be:
			% right,left,above,below,
			% above left,below right, etc
			\node[state] (y) [below right =of z] {$V_2$};
			
			% y node set relative to x.
			% Locations can be:
			% right,left,above,below,
			% above left,below right, etc
			\node[state] (x) [below left=of z] {$V_1$};
			
			% Directed edge
			\path (x) edge (z);
			
			% Directed edge
			\path (y) edge (z);
			
			% Bidirected edge
			%\path[bidirected] (x) edge[bend left=60] (y);
		\end{tikzpicture}
		
	\end{minipage}
	\begin{minipage}{.25\paperwidth}
		
		\begin{tikzpicture}
			% x node set with absolute coordinates
			\node[state] (z) at (0,0) {$V_3$};
			
			% y node set relative to x.
			% Locations can be:
			% right,left,above,below,
			% above left,below right, etc
			\node[state] (y) [below right =of z] {$V_2$};
			
			% y node set relative to x.
			% Locations can be:
			% right,left,above,below,
			% above left,below right, etc
			\node[state] (x) [below left=of z] {$V_1$};
			
			% Directed edge
			\path (x) edge (y);
			
			% Directed edge
			\path (x) edge (z);
			
			% Directed edge
			\path (z) edge (y);
			
			% Bidirected edge
			%\path[bidirected] (x) edge[bend left=60] (y);
		\end{tikzpicture}
		
	\end{minipage}
	
\end{center}

\vspace{\baselineskip}

Further suppose that, in the graph on the right, the causal influence of V$_1$ on V$_2$ is precisely canceled by the causal influence of V$_3$ on V$_2$. Then both of these graphs entail V$_1 \indep \text{V}_2$ and V$_1 \slashed{\indep} \text{V}_2 \, \mid \text{V}_3$, but there is something unsatisfying about the graph on the right: the required conditional dependence relations have been recovered only by fine-tuning causal influences to cancel precisely. This is why \citep[chapter 2.4]{2009_pearl} calls this a stability condition: holding fixed the strength of the causal influence V$_1$ on V$_2$ while perturbing the strength of the causal influence of V$_3$ on V$_2$ by any $\delta > 0$ will destroy the conditional independence relations. It is these kind of finely-tuned graphs that are ruled out by faithfulness.\footnote{Faithfulness is often motivated by the fact that the set of parameter values quantifying causal influence that produce this type of cancellation are measure zero in the set of all parameter values \cite{1995_meek}, \cite[theorem 3.2]{2000_spirtes_glymour_scheines}. For a clarifying discussion of alternative justifications for imposing faithfulness, see \citep{2018_weinberger}.}

In short, if a probability distribution satisfies the Causal Markov and Faithfulness conditions relative to a candidate causal graph that entails all, and only, the conditional dependence relations present in the joint distribution P(V$_1$, V$_2$, $\ldots$, V$_\text{N}$) are reflected in the causal structure of the graph. If a joint probability distribution satisfies these two properties relative to multiple graphs, those graphs are said to form a \textit{Markov equivalent} set.

One can now see why these conditions are insufficient for causal inference with two variables V$_1$ and V$_2$. Suppose one already knows that any statistical dependence between the two variables is due to a direct causal relationship between them, allowing them to rule out confounders. Even then, the above conditions are insufficient to identify whether V$_1$ causes V$_2$ or vice versa. One is stuck with the following Markov equivalence class of graphs: 

\vspace{\baselineskip}

\begin{center}
	
	\begin{minipage}{.25\paperwidth}

		\begin{tikzpicture}
			% x node set with absolute coordinates
			\node[state] (z) at (0,0) {$V_1$};
			
			% y node set relative to x.
			% Locations can be:
			% right,left,above,below,
			% above left,below right, etc
			\node[state] (y) [right =of z] {$V_2$};
			
			% y node set relative to x.
			% Locations can be:
			% right,left,above,below,
			% above left,below right, etc
			%\node[state] (x) [below left=of z] {$V_1$};
			
			% Directed edge
			\path (z) edge (y);
			
			% Directed edge
			%\path (y) edge (z);
			
			% Bidirected edge
			%\path[bidirected] (x) edge[bend left=60] (y);
		\end{tikzpicture}
		
	\end{minipage}
	\begin{minipage}{.25\paperwidth}
		
		\begin{tikzpicture}
			% x node set with absolute coordinates
			\node[state] (z) at (0,0) {$V_2$};
			
			% y node set relative to x.
			% Locations can be:
			% right,left,above,below,
			% above left,below right, etc
			\node[state] (y) [left =of z] {$V_1$};
			
			% y node set relative to x.
			% Locations can be:
			% right,left,above,below,
			% above left,below right, etc
			%\node[state] (x) [below left=of z] {$V_1$};
			
			% Directed edge
			\path (z) edge (y);
			
			% Directed edge
			%\path (y) edge (z);
			
			% Bidirected edge
			%\path[bidirected] (x) edge[bend left=60] (y);
		\end{tikzpicture}
		
	\end{minipage}
	
\end{center}

\vspace{\baselineskip}

\noindent The reason is simple: any causal graph relating the two variables V$_1$ and V$_2$ will be fully connected, and fully connected graphs do not predict any conditional independence relations. This is why causal inference from conditional  independences in observational statistical data, using only the Causal Markov Property and Faithfulness, can successfully identify the causal relationship between two variables only if they are embedded in a larger set of at least three variables.
% Learning information about the temporal ordering of and Y can often be sufficient, but that is precisely what the time-symmetry of quantum theories prevents us from determining solely from the knowledge that $\ket{\alpha}$ and $\ket{\beta}$ are related by some unitary time evolution $U(t)$. 

Happily, it turns out that one can do better than this if they are willing to introduce additional conditions on a satisfactory causal graph. A set of causal inference strategies has recently been developed that aims to address the problem of causal inference from observational data for only two variables \citep{2008_janzing_scholkopf,2010_daniusis_janzing_mooij,2012_daniusis_janzing,2017_peters_janzing_scholkopf}. The central condition they introduce is simple: given two candidate causal graphs that generate the joint distribution P(V$_1$, V$_2$), the true causal graph is the one that entails that the distribution for the ``cause'' variable P(V$_1$) does not contain any information about the conditional distribution P(V$_2 \mid \text{V}_1$) for the ``effect'' variable, and vice versa. A perhaps more intuitive way to describe this condition is as follows: the mechanism that determines the probability distribution over the cause variable operates independently of the mechanism that determines the conditional distribution over the effect variable, given the distribution over the cause; \citep[chapter 2]{2017_peters_janzing_scholkopf} label this ``The Principle of Independent Mechanisms''. 

\begin{wrapfigure}{r}{0.4\textwidth} %this figure will be at the right
	\centering
	\includegraphics[width=0.4\textwidth]{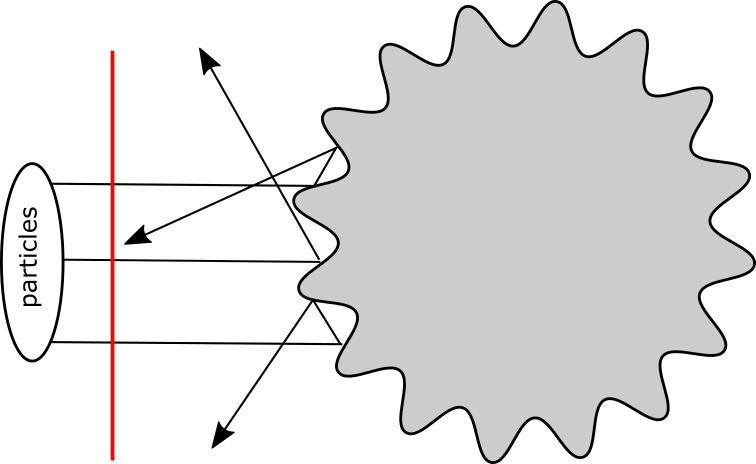}
	\caption{A jagged potential.}
\end{wrapfigure}

The range of contexts in which such a principle is justified has been given an enlightening examination in \citep{2020_woodward}, but is quite reasonable in the particular context of physical theory: one can specify an initial ``cause'' state -- or a probability distribution over initial ``cause'' states -- without any detailed knowledge of the dynamical mechanism that will take that initial state as input and output a conditional distribution over final, or ``effect'', states. If that is possible according to the way P(V$_1$, V$_2$) factorizes according to one candidate causal graph but impossible according to the second, then it seems quite reasonable that the first causal graph is the correct one. 

One can illustrate the principle with a simple example. Suppose one prepares a beam of \textit{N} classical particles and scatters them off a jagged potential like the one in Figure 1.\footnote{This is roughly a combination of two different examples from \citep[chapter 2]{2017_peters_janzing_scholkopf} and \citep{2016_janzing_chaves_scholkopf}.} Draw a line that all \textit{N} particles will cross, both before they encounter the potential \textit{and} after they are scattered back off it, some distance away from the region where the potential is non-zero. Repeat the experiment many times and record the position of the particles each time they cross the line on their way in and again on their way out. This will produce a joint probability distribution over the positions of the N particles P(X$_{\text{in}}$, X$_{\text{out}}$), and the scattering potential is the mechanism that determines the conditional distributions P(X$_{\text{in}} \mid \text{X}_{\text{out}})$ and P(X$_{\text{out}} \mid \text{X}_{\text{in}})$. The causal inference problem is whether one could infer that one distribution is the cause and the other the effect \textit{in the absence of any temporal information}. 
		
The Causal Markov and Faithfulness conditions are insufficient, as discussed above, but it is easy to see how the Principle of Independent Mechanisms can make this possible. The distribution P(X$_\text{out}$) will be highly disordered, with the positions of the \textit{N} particles distributed roughly randomly across the line. However P(X$_\text{in}$) will be quite uniform: the beam of \textit{N} particles will cross the line at roughly the same positions on each run of the experiment. For P(X$_\text{out}$) to be the cause distribution, it would have to be such that the potential would funnel the random distribution into a highly ordered one. This would be impossibly unlikely given the geometry of the scattering potential unless each state of the incoming beams of \textit{N} particles had been \textit{extremely} carefully fine-tuned to be funnelled into a more ordered state by that particular scattering potential. Any such preparation procedure would require extremely detailed information about the geometry of the mechanism determining P(X$_{\text{in}} \mid X_{\text{out}})$ -- the scattering potential -- and so identifying P(X$_\text{out}$) as the cause would violate the Principle of Independent Mechanisms. 
			
A natural precisification of this principle uses algorithmic information theory: the true causal graph should render the \textit{mutual information} $\text{I}(P_{\text{cause}} \, : \, P_{\text{effect}\mid\text{cause}})$ between the probability distribution over the cause and the conditional distribution over the effect equal to zero. This can be stated in an equivalent form that is more transparent in the context of physical theory: the true causal graph should render equal to zero the mutual information $\text{I}(s \, : \, M)$ between the ``cause'' state \textit{s} and the dynamical mechanism \textit{M} that determines the conditional distribution over the effect, given \textit{s}. In fact, Janzing, Sch\"olkopf, and collaborators have constructed a framework for causal inference that is founded on the use of the tools of algorithmic information theory in which The Principle of Indepenent Mechanisms plays a foundational role.

A valuable feature of this framework is that it is \textit{non-probabilistic}: its foundational concepts are those of algorithmic information theory, not statistics \citep{2008_janzing_scholkopf,2017_peters_janzing_scholkopf}. This is not to say that it somehow doesn't work for the more familiar cases of causal inference from statistical data; in fact, the examination of algorithmic informational dependencies can reveal novel statistical dependencies entailed by a causal graph \citep[section 3]{2008_janzing_scholkopf}. The point is that it also applies more generally, enabling causal inference between objects even when one does not have statistical data about those objects, like a joint probability distribution over their values. Although I do not rely on it here, this may prove valuable for doing causal inference in quantum theory where joint probability distributions are not guaranteed to be well-defined (see fn. \ref{fnjointdist}).
			
I will begin with a brief introduction to some relevant concepts of algorithmic information theory.\footnote{I very loosely follow \citep[section 2.1]{2008_janzing_scholkopf} here. See \citep{2019_li_vitanyi} for a textbook introduction.} Suppose one has a countable set of objects $\left\{ \mathcal{O}_1, \, \mathcal{O}_2, \, \ldots, \, \mathcal{O}_n, \, \ldots \right\}$ and a system for identifying each object $\mathcal{O}_i$ by a binary description $s_i$. Note that ``objects'' here is quite broad: it can include things like probability distributions, states of a physical system, PDF files, \textit{Blood Meridian}, Gila monsters, surfboards, etc. The algorithmic information content of an object $\mathcal{O}_i$ is meant to quantify the amount of information required by the \textit{shortest} complete description of  the object $\mathcal{O}_i$.  

There will always be some shortest binary string that uniquely identifies an object $\mathcal{O}$, denoted $s^*$. The algorithmic information of $\mathcal{O}$ (also called the algorithmic complexity or Kolmogorov complexity) is the length $\ell (p)$ of the shortest program \textit{p} that, if run on a (prefix-free) universal Turing machine \textit{U}, would output the string $s^*$ and halt. Formally:\footnote{In algorithmic information theory, equalities are generally only equalities up to a constant that is independent of the object itself, but may depend on the alphabet or programming language chosen for the encoding or the particular universal Turing machine being considered. For example, a program to output \textit{Hamlet} may be shorter when written in Python than in FORTRAN. This tells us something about Python and FORTRAN, but nothing about the algorithmic information content of \textit{Hamlet} itself. This ``equality up to a constant'' is denoted by $\stackrel{+}{=}$}
		
	\[
	K(s) = \min_{p} \left\{\ell (p) \mid U(p) = s \right\}
	\]
		
\noindent Note that the upper bound on $K(s)$ for an \textit{n}-bit string is \textit{n} since one can always write a program of the form ``Print $s_1, \, \ldots , \, s_n$'' that reproduces the string $K(s)$ bit-by-bit. This illustrates one sense in which $K(s)$ is a measure of the information contained in the object $\mathcal{O}$: it captures how much information is required to algorithmically reconstruct its complete binary description.\footnote{Note that this has the somewhat counterintuitive consequence that a binary sequence that is completely random has maximal algorithmic information.} 
		
One can similarly define the conditional algorithmic information of one object, given a second. Let $t$ and $s$ be the shortest binary descriptions of objects $\mathcal{O}_t$ and $\mathcal{O}_s$. Then the conditional algorithmic information of $\mathcal{O}_s$ given $\mathcal{O}_t$, denoted $K(s\mid t)$, is defined as as the length of the shortest program that takes as input \textit{t} then generates \textit{s} as output and halts. This measures how much information about $\mathcal{O}_s$ one obtains if given $\mathcal{O}_t$, and thus how much computational work is saved by knowing \textit{t} when computing \textit{s}. If \textit{t} contains no information about \textit{s} then $K(s\mid t) = K(s)$.
		
The conditional algorithmic information allows a precise definition of the natural intuition that a cause does not contain information about the mechanism that maps it to the effect. The \textit{mutual algorithmic information} between two objects $\mathcal{O}_t$ and $\mathcal{O}_s$ is:
		
\[
	I(s \, : \, t) = K(t) - K(t\mid s^*) 
\]

\noindent One can also use the \textit{joint algorithmic information} $K(s, t) \stackrel{+}{=} K(s) + K(t \mid s^*)$ to give a symmetric formulation of the mutual algorithmic information:

\[
I(s \, : \, t) \stackrel{+}{=} K(s) + K(t) - K(s ,t)
\]
		
\noindent This gives the intuitive result that if knowing $s^*$ does not enable a shorter computation of \textit{t} (or vice-versa) then \textit{s} and \textit{t} do not share any mutual information. 

In \citep[section 2]{2008_janzing_scholkopf} they used a related concept, the conditional mutual information

\[
I(s \, : \, t \mid r) \stackrel{+}{=} K(s \mid r) + K(t \mid r) - K(s, t \mid r)
\]

\noindent to define an algorithmic information theoretic formulation of the Causal Markov Property. If $x_1, \, x_2, \, \ldots, \, x_n$ are binary strings describing observations related by a graph, then each string $x_j$ shares no mutual information with the strings $nd_j$ associated with its non-descendents in the graph, conditional on the \textit{shortest} binary string associated with the parents $pa_j^*$ of $x_j$ in the graph:

\[
I(x_j \, : \, nd_j \mid pa_j^*) \stackrel{+}{=} 0
\]

\noindent In general, Janzing, Sch\"olkopf, and collaborators have made use of algorithmic information theory for a multitude of causal inference tasks; see \citep{2016_mooji_peters_janzing} for a review of how some of these methods perform on a variety of empirical data sets.

My interest is in the application of some of these tools to inferring causal direction in physical theory. In particular, I will focus on a connection drawn in \citep{2016_janzing_chaves_scholkopf} between dynamical evolution and the increase of algorithmic complexity in a toy model of statistical mechanics, with the aim of extending it to quantum theories.
		
They begin by assuming the Principle of Independent Mechanisms: if \textit{s} is the initial state of an \textit{N}-particle system and the operator $D_t$ represents applying the dynamics governing the system for some time interval \textit{t}, then $I(s \, : \, D_t) \stackrel{+}{=} 0$. (More precisely, \textit{s} and $D_t$ are the shortest binary encodings of the initial state and the dynamics.) This asserts that not only does \textit{s} contain no information about the dynamics \textit{D} (the geometry and strength of a potential, for example), \textit{s} also contains no information about how much time it will be subjected to those dynamics (i.e. it contains no information about the interval \textit{t}). 
		
Although it seems so weak as to be nearly tautological, Janzing, Chaves, and Sch\"olkopf immediately derive from it that the \textit{algorithmic entropy} $K(s)$ of a toy \textit{N}-particle system \textit{must} increase under dynamical evolution.\footnote{The equation of algorithmic entropy with $K(s)$ by \citep{2016_janzing_chaves_scholkopf} assumes that the microstate \textit{s} is perfectly known to the observer. More generally, one defines the algorithmic entropy of a microstate \textit{s} as the sum of the algorithmic information and the thermodynamic entropy $\mathcal{S}(s) = K(s) + H(s)$; see \citep{1989_zurek} or \citep[chapter 8]{2019_li_vitanyi}.} The proof is sufficiently simple that I will include it here before flagging its main limitation:
		
\vspace{\baselineskip}
		
\noindent \textbf{No entropy decrease}: \textit{If the dynamics of a system is an invertible mapping $D_t$ of a finite set $\mathcal{S}$ of states then $I(s \, : \, D_t) \stackrel{+}{=} 0$ implies that the algorithmic information can never decrease when applying $D_t$ to the initial state, i.e.}
		
	\[
	K(D(s)) \stackrel{+}{\geq} K(s)
	\]
		
\noindent \textit{for all $s \in \mathcal{S}$}.

\vspace{\baselineskip}
		
\noindent Proof: Imposing $I(s \, : \, D_t) \stackrel{+}{=} 0$ entails that $K(s) \stackrel{+}{=} K(s \mid D_t)$. Since $D_t$ is invertible, \textit{s} can be computed from $D_t (s)$ and vice versa, which implies $K(s \mid D_t) \stackrel{+}{=} K(D_t (s) \mid D_t)$. From this one has $K(s) \stackrel{+}{=} K(s \mid D_t) \stackrel{+}{=} K(D_t (s) \mid D_t) \stackrel{+}{\leq} K(D_t (s))$. \hfill $\square$

\vspace{\baselineskip}
		
\noindent The intuitive idea is simple: if $D_t (s)$ had a shorter description than \textit{s} one could obtain a shorter binary description of the initial state by encoding $D_t (s)$ and adding ``then apply  $D_t^{-1}$''. But that is impossible since we've assumed \textit{s} is the shortest binary description of the initial state. Upon establishing this theorem they illustrate that it holds for a toy system of N particles modeled as a cellular automaton \citep[section 2]{2016_janzing_chaves_scholkopf}.
		
The theorem is suggestive but, as it stands, ultimately insufficient for inferring causal direction for time evolution in quantum systems. The reason for this is simple: the set of possible states of any quantum system is infinite. That said, I showed by construction at the end of section 3 that something much like \textbf{No entropy decrease} should be true for quantum systems: the Schmidt rank of a bipartite quantum system in a pure state is guaranteed to not decrease under time evolution, except for a set of time intervals $t_i$ of measure zero, because time evolution generically creates entanglement.
		
To connect my discussion at the end of section 3 to the methods of causal inference using algorithmic information theory, I need a measure of the algorithmic information of a quantum state. In the next section, I will show that one can use such a measure and the causal inference methods described above to derive a more general, and more useful, version of the conclusion I reached at the end of section 3.

\section{Entanglement, Algorithmic Information, and Causal Direction}

To embed the inference of causal direction made on physical grounds in section 3 into the framework of causal inference using algorithmic information theory, one needs a measure of the algorithmic information of a quantum state. It would be preferable for the measure to be natural, either in the sense that it shares many or all of the conceptual virtues of classical algorithmic complexity or because it seems to appropriately generalize those properties to the novel physical and mathematical setting of quantum theories. Ideally, there would be a unique such measure. 

There are multiple proposals for quantitative meaures of quantum algorithmic information, each with some legitimate claim to being a natural generalization of classical algorithmic information to quantum states (see \citep{2001_vitanyi} for an early overview and \citep{2007_mora_briegel_kraus} for a more recent one). We do not live in the best of all possible worlds, however: the measures are demonstrably inequivalent and one thus has to make a choice. I will focus on a measure of algorithmic information that tracks the Schmidt rank of a quantum state (or the Schmidt measure, for multipartite systems). Ultimately I will show that one can reproduce a less restrictive version of the criteria for identifying causal direction offered at the end of section 3 as a theorem that follows from imposing the Principle of Independent Mechanisms. I will first say a bit about the problems faced by any extension of algorithmic information to quantum states.

The question of how much a measure of quantum algorithmic information needs to have in common with its classical precursor to deserve the name is somewhat subjective, but one can identify at least three properties that seem non-negotiable. First, it should be definable from the quantum state alone rather than, say, only for a member of a specified ensemble of quantum states. This was why I deemed the von Neumann entropy unsatisfactory in section 3. Second, there should be some recognizable sense in which it measures the amount of information required to compute, or reconstruct, the state in question using some algorithmic procedure. Finally, just as there is an upper bound on the algorithmic information required to specify any classical bit string of specified length \textit{n}, there should be some analogous upper bound on the algorithmic information of a quantum state that scales with the size of the system.

A number of other seemingly essential properties of classical algorithmic information are up for grabs, however. Must quantum algorithmic information be defined in terms of a computation carried out by a quantum generalization of a Turing machine or is a different notion of ``algorithmic procedure'' appropriate? Should the algorithmic information be measured in classical bits or qubits? Does quantum algorithmic information need to reduce smoothly to classical information in some context? \textit{How} should the upper bound on the algorithmic information of a quantum system scale with the size of the system? Almost all classical bit strings of length \textit{n} maximize the classical algorithmic information; should the same be true of the quantum algorithmic information content of quantum states in, say, an \textit{N}-dimensional Hilbert space? Different proposals for measures of quantum algorithmic complexity have adopted different answers to these questions \citep{2001_berthiaume_van_dam_laplante,2001_vitanyi,2007_mueller}.

The measure I will focus on defines the amount of algorithmic information in a quantum state $\ket{\phi}$ as the \textit{classical} algorithmic information required by the shortest description of an algorithmic procedure for preparing it from some reference state $\ket{0}$ \citep{2005_mora_briegel,2006_mora_briegel}. One models the algorithmic preparation procedure as a quantum circut $\mathcal{C}_\phi$: a finite sequence of unitary operations -- quantum gates -- chosen from a finite set $\mathcal{G} = \left\{G_1, \, \ldots , \, G_n \right\}$ that, when performed on the reference state $\ket{0}$, produce the desired state $\mathcal{C}_\phi \ket{0} = \ket{\phi}$ up to some fidelity $\varepsilon$.\footnote{This means that $\abs{\matrixel{\phi}{\mathcal{C}}{0}}^2 \geq 1 - \varepsilon$.} The set $\mathcal{G}$ is finite, so each unitary operation -- and thus each circuit $\mathcal{C}_\phi$ -- can be given a finite binary encoding which has a finite quantity of classical algorithmic information. The algorithmic information of a quantum state $\ket{\phi}$ is identified with the classical algorithmic information of the binary encoding of the simplest circuit $\mathcal{C}_\phi$:

\[
K^\varepsilon_Q (\ket{\phi}) = \min_{\mathcal{C}_\phi} K (\mathcal{C}_\phi)
\]

\noindent where the notation $\mathcal{C}_{\phi}$ is doing double-duty as the quantum circuit or the binary encoding of that circuit, depending on context. 

The value of $K^\varepsilon_Q$ appears to depend on four quantities not obviously related to the state itself: (i) the choice of the set $\mathcal{G}$ of quantum gates, (ii) the alphabet $\Omega$ used to encode the circuit, (iii) the degree of fidelity $\varepsilon$, and (iv) the particular circuit $\mathcal{C}_{\phi}$. As \citep{2005_mora_briegel,2006_mora_briegel} show, (ii) and (iv) are unproblematic: the values of $K^\varepsilon_Q (\ket{\phi})$ for two different alphabets $\Omega_1$ and $\Omega_2$ differ by at most a constant, and minimizing over all circuits $\mathcal{C}_{\phi}$ that prepare the state $\ket{\phi}$ removes any dependence on an arbitrary choice of circuit. On reflection, it is physically quite reasonable that $K^\varepsilon_Q$ depend on the degree of fidelity $\varepsilon$: changing the precision with which an algorithm has to prepare a state will, and should, change the amount of information the algorithm requires. The arbitrariness in $K^\varepsilon_Q$ associated with (i), the chosen set of quantum gates $\mathcal{G}$, is genuine, although \citep{2005_mora_briegel,2006_mora_briegel} offer several palliative remarks on that front.

To see the connection between this notion of algorithmic information and entanglement, consider a quantum system consisting of \textit{N} qubits with Hilbert space $\mathcal{H}_N$. In this case, one can derive the specific dependence of $K^\varepsilon_Q$ on the fidelity $\varepsilon$ by invoking the fact that beginning with the reference state $\ket{0} = \ket{0}_1 \, \cdots \, \ket{0}_N \in \mathcal{H}_N$, one can prepare any state $\ket{\phi} \in \mathcal{H}_N$ with \textit{at most} $\order{2^N}$ quantum gates $G_i$. Taking into account the dependence on the fidelity $\varepsilon$, \citep[section 5]{2006_mora_briegel} show that the number of quantum gates required to compose a circuit $\mathcal{C}_\phi$ that prepares an arbitrary state $\ket{\phi}$ from the reference state $\ket{0}$ is $\mathcal{O}(2^N\log \frac{1}{\varepsilon})$. This is an intuitive result: the bigger the system or the greater fidelity demanded, the more unitary operations required.

\begin{wrapfigure}{r}{0.4\textwidth} %this figure will be at the right
	\centering
	\includegraphics[width=0.3\textwidth]{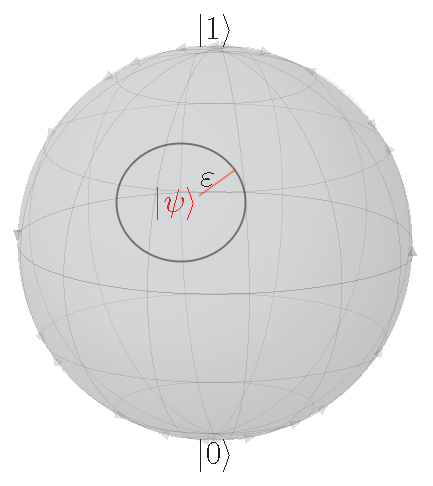}
	\caption{An $\varepsilon$-radius patch around $\ket{\psi}$.}
\end{wrapfigure}

Note that by specifying a degree of fidelity $\varepsilon$ one identifies a \textit{patch} around $\ket{\phi}$. More precisely, it specifies a patch of radius $\varepsilon$ on the unit sphere corresponding to all pure states in $\mathcal{H}_N$; see Figure 2. The quantity $K^\varepsilon_Q(\ket{\phi})$ captures the amount of information required to specify a circuit $\mathcal{C}_\phi$ that will put $\ket{0}$ in that patch.\footnote{The two limiting cases may be clarifying. Let $\varepsilon \to 1$; then the ``patch'' is the entire sphere of pure states and $K^{\varepsilon=1}_Q(\ket{\phi})$ quantifies the information required to construct a circuit $\mathcal{C}$ that will act on $\ket{0}$ to put it into \textit{any} pure state inside $\mathcal{H}_N$. In this limit $K^{\varepsilon=1}_Q$ is zero for all states, as it obviously should be: $\ket{0}$ is already such a state, so just leave it alone. Now let $\varepsilon \to 0$; then the ``patch'' approaches the single state $\ket{\phi}$ and $K^{\varepsilon=0}_Q(\ket{\phi})$ quantifies the information required to prepare \textit{exactly} $\ket{\phi}$ from $\ket{0}$. This obviously diverges at the limit; for example, it would require preparing an amplitude to be \textit{exactly} $\frac{1}{\sqrt{2}}$ rather than the closest rational approximation to $\frac{1}{\sqrt{2}}$.} Choosing a degree of fidelity $\varepsilon$ thus defines an equivalence class of states close to $\ket{\phi}$ by assigning each of them the same algorithmic information content $K^\varepsilon_Q(\ket{\phi})$. For $\varepsilon > 0$ such ``patches'' of states have small but finite measure in the set of pure states in $\mathcal{H}_N$; indeed, one can cover the unit sphere of pure states with finitely many of them\footnote{See \citep[section 4.5.4]{2010_nielsen_chuang} for how the number of patches required scales with the fidelity $\varepsilon$ and system size \textit{N}.} Topologically, this means that every state $\ket{\phi}$ is surrounded by a ball of radius $\varepsilon$ of states of equivalent algorithmic information.
	
The information required to describe the circuit $\mathcal{C}_\phi$ is just the information required to describe the quantum gates that compose it. From the fact that $\mathcal{C}_\phi$ consists of, at most, a sequence of $\sim 2^N \log \frac{1}{\varepsilon}$ unitary operations, \citep{2006_mora_briegel} conclude that its binary encoding -- and thus the quantum algorithmic information of the state $\ket{\phi}$ -- has the upper bound

\[
K^\varepsilon_Q (\ket{\phi}) = \min_{\mathcal{C}_\phi} K (\mathcal{C}_{\phi}) \lesssim 2^N \log \frac{1}{\varepsilon}
\]

\noindent It is significant that while the classical algorithmic information of an \textit{n}-bit string grows at most linearly with \textit{n}, the algorithmic information of a quantum state can grow \textit{exponentially} with the size of the system \textit{N}. Given that the dimension of the space of states of a composite classical system in physics generally grows linearly in the dimension of the subsystems, while the dimension of the space of states of a composite quantum system grows exponentially due to the possibility of entanglement, one might think that the upper bound of $K^\varepsilon_Q (\ket{\phi})$ reflects the presence of entangled states in $\mathcal{H}_N$. As \citep{2005_mora_briegel,2006_mora_briegel} show, that is correct.

At this point the Schmidt rank of a quantum state once again becomes relevant. They adopt as their measure of entanglement the Schmidt measure \citep{2001_eisert_briegel}, a generalization of the Schmidt rank to multipartite quantum systems. For an \textit{N}-particle system with Hilbert space $\mathcal{H} = \mathcal{H}_1 \otimes \cdots \otimes \mathcal{H}_N$, any state $\ket{\Psi}$ in $\mathcal{H}$ can be written as\footnote{I am presuming a preferred factorization of the big Hilbert space $\mathcal{H}$ into tensor factors. In general, entanglement measures for multipartite systems are sensitive to different choices of partition; this is part of the difficulty of extending such measures beyond bipartite systems. Where a preferred partition isn't available, one could more generally define the Schmidt measure as the minimum value of \textit{r} over all possible partitions. See \citep{2004_zanardi_lidar_lloyd}, \citep{2019_cotler_penington_renard}, or \citep{2020_carroll_singh} for different proposals for identifying preferred factorizations in certain contexts.}

\[
\ket{\Psi} = \sum^R_i a_i \ket{\psi_i^{(1)}} \otimes \cdots \otimes \ket{\psi_i^{(N)}}
\]

\noindent Let \textit{r} be the minimum number of product states $\ket{\psi^{(1)}} \otimes \cdots \otimes \ket{\psi^{(N)}}$ needed to write $\ket{\Psi}$ in the above form. The \textit{Schmidt measure} of the state $\ket{\Psi}$ is defined as $S(\ket{\Psi}) = \log r$. A separable state will have Schmidt measure $\log 1 = 0$ and a fully entangled state will have Schmidt measure $\log \dim (\mathcal{H})$, which will be $\log 2^N = N \log 2 = N$ for the system of \textit{N} qubits.%\footnote{Like the Schmidt rank, maximal entanglement according to the Schmidt measure will typically not agree with maximal entanglement as defined by measures like the entanglement entropy.}

The connection between algorithmic information and entanglement follows from the fact one can use the Schmidt measure to bound the algorithmic information $K^\varepsilon_Q$ of a quantum state. As \citep{2006_mora_briegel} show, knowing the Schmidt measure of a quantum state makes possible a more informative bound on its complexity $K^\varepsilon_Q$:

\[
K^\varepsilon_Q (\ket{\phi}) = \min_{\mathcal{C}_\phi} K (\mathcal{C}_{\phi}) \lesssim 3N2^{S(\ket{\phi})}\log \frac{1}{\varepsilon}
\]

\noindent The algorithmic complexity of a separable state $\ket{1}$ can grow at most linearly with $N$, while the upper bound for all states in $\mathcal{H}_N$ is saturated by a state $\ket{N}$ only if it has full Schmidt measure. Given a state $\ket{k}$ that is neither separable nor fully entangled -- i.e. a state which has Schmidt measure $0 < k < N$ --  the upper bound on the algorithmic complexity of $\ket{k}$ will thus satisfy $K^\varepsilon_Q (\ket{\text{SEP}}) < K^\varepsilon_Q (\ket{k}) < K^\varepsilon_Q (\ket{N})$.

%\vspace{\baselineskip}

%\begin{center}
	
%\begin{blochsphere}[radius=2.5 cm,tilt=15,rotation=-20,opacity=0.25]
%	\drawBallGrid[style={opacity=0.15}]{30}{30}
	
	%\drawGreatCircle[style={dashed}]{0}{0}{0}
	%\drawGreatCircle[style={dashed}]{90}{0}{0}
	
	%\drawRotationLeft[scale=1.3,style={red}]{-60}{0}{0}{15}
	%\drawRotationRight[scale=1.3,style={red}]{60}{0}{0}{15}
	
%	\node at (-0.5,0.5) {\textcolor{red}{ $\ket{\psi}$}};
%	\node at (-0.1,0.8) {\textcolor{black}{$\varepsilon$}};
	
	%\drawSmallCircle [] {30} {30} {0.75}
	
%	\labelLatLon{up}{90}{0};
%	\labelLatLon{down}{-90}{90};
%	\node[above] at (up) {{$\ket{1}$ }};
%	\node[below] at (down) {{$\ket{0}$}};
	
	%\labelLatLon[labelmark=false]{d}{15}{90};
	%\node at (d) {\color{gray}\fontsize{0.15cm}{1em}\selectfont $60^\circ$};
	
	%\labelLatLon[labelmark=false]{d2}{5}{78};
	%\node at (d2) {\color{gray}\fontsize{0.15cm}{1em}\selectfont $60^\circ$};
%\end{blochsphere}

%\end{center}

There are several comforting features of this definition of quantum algorithmic information. The first is that it lays the groundwork for addressing two major limitations of the causal inference criteria of section 3: that they applied only if one state from the pair comes from a set of states that has measure zero in the set of all pure states in $\mathcal{H}_{AB}$, and that they no longer held if one replaced states $\ket{\alpha}$ that were \textit{exactly} of less-than-full Schmidt rank with states that were \textit{within $\varepsilon$} of a state of less-than-full Schmidt rank. In algorithmic information theoretic terms, states $\ket{k}$ of less-than-full Schmidt rank now correspond to $\varepsilon$-\textit{patches} of non-maximal algorithmic information $K^\varepsilon_Q (\ket{k})$. As mentioned above, those patches have non-zero measure on the sphere of pure states in $\mathcal{H}_N$. Furthermore, these patches include all states that \textit{approximate} $\ket{k}$ with fidelity $\varepsilon$. According to Schmidt rank alone, the two states 

\[
\ket{\Psi} = \ket{00} \hspace{24pt} \ket{\Psi_\varepsilon} = \sqrt{1-\varepsilon}\ket{00} + \varepsilon\ket{11}
\]  

\noindent are no less different than are

\[
\ket{\Psi} = \ket{00} \hspace{24pt} \ket{\Psi_\text{EPR}} = \frac{1}{\sqrt{2}}\ket{00} + \frac{1}{\sqrt{2}}\ket{11}
\]

\noindent The algorithmic information $K^\varepsilon_Q$ is less sensitive: it treats $\ket{\Psi}$ and $\ket{\Psi_\varepsilon}$ as equivalent, but distinguishes $\ket{\Psi}$ from $\ket{\Psi_\text{EPR}}$. It represents an improvement over the Schmidt rank in this regard.

Nevertheless it can still reproduce an algorithmic information-theoretic version of the topological and measure-theoretic facts relied upon in section 3. In \citep{2006_mora_briegel} they show that for a system of \textit{N} qubits, the relative frequency of pure states in $\mathcal{H}_N$ with non-maximal algorithmic information is exponentially small: it scales as $\sim 2^{2^N \log \varepsilon}$.\footnote{Since the algorithmic information of a quantum state is the classical algorithmic information of the circuit $\mathcal{C}$ that prepares it, this is a straightforward adaptation of the standard proof that the relative frequency of compressible strings in the set of all \textit{n}-bit strings goes to zero as $n\to\infty$.} Recall that (i) a state in $\mathcal{H}_N$ is maximally complex only if it has full Schmidt measure and (ii) the Schmidt measure reduces to the Schmidt rank for pure states of a bipartite system. With that in mind, consideration of the (divergent) limit $\varepsilon \to 0$ shows that the relative frequency of pure states of non-maximal algorithmic information goes to zero in the set of all pure states as one demands perfect fidelity. One can thus reproduce in algorithmic information-theoretic terms the measure-theoretic sparseness of states that are not fully entangled, not only for bipartite pure states but also for pure states of \textit{multipartite} systems of \textit{N} qubits.
%\footnote{If one is unsettled by the fact that $K^\varepsilon_Q$ diverges at the limit $\varepsilon = 0$, one can reach the same conclusion -- at least, for bipartite systems -- by observing that the volume of the ``patches'' defined by a choice of fidelity $\varepsilon$ will vanish as $\varepsilon \to 0$, and the ball of radius $\varepsilon$ of states of equal algorithmic information around each pure state will shrink to a point: that state itself. One thus returns to the set of pure states in $\mathcal{H}_{AB}$ and recovers the topological and measure-theoretic sparseness of non-maximally entangled states reviewed in section 3.}

With all this machinery in place, an analogue of the result of \citep{2016_janzing_chaves_scholkopf} for quantum algorithmic information now follows by essentially identical reasoning. 

\vspace{\baselineskip}

\noindent \textbf{No entanglement decrease}: \textit{If the dynamics of a quantum system are given by a unitary operator $U_t$ on a finite-dimensional Hilbert space $\mathcal{H}$ of states $\ket{\phi}$, then $I(\ket{\phi} \, : \, U_t) \stackrel{+}{=} 0$ implies that the quantum algorithmic information can never decrease when applying $U_t$ to the initial state, i.e.}

\[
K^\varepsilon_Q(U_t \ket{\phi}) \stackrel{+}{\geq} K^\varepsilon_Q( \ket{\phi})
\]

\noindent \textit{for all $\ket{\phi} \in \mathcal{H}$}.

\vspace{\baselineskip}

\noindent Proof: The assumption that $I(\ket{\phi} \, : \, U_t) \stackrel{+}{=} 0$ entails that $K^\varepsilon_Q(\ket{\phi}) \stackrel{+}{=} K^\varepsilon_Q(\ket{\phi} \mid U_t)$. Unitary time evolution is invertible, so $\ket{\phi}$ can be computed from $U_t \ket{\phi}$ and vice versa. This implies that $K^\varepsilon_Q(\ket{\phi} \mid U_t) \stackrel{+}{=} K^\varepsilon_Q(U_t \ket{\phi} \mid U_t)$. From this, one obtains $K^\varepsilon_Q(\ket{\phi}) \stackrel{+}{=} K^\varepsilon_Q(\ket{\phi} \mid U_t) \stackrel{+}{=} K^\varepsilon_Q(U_t \ket{\phi}\mid U_t) \stackrel{+}{\leq} K^\varepsilon_Q(U_t \ket{\phi})$. \hfill $\square$

\vspace{\baselineskip}

\noindent This is unsurprising  in light of  the facts reviewed above concerning the relative (in)frequency of states of non-maximal complexity in the set of all pure states. 

One can also recognize the same intuitive reasoning at work as in the result of \citep{2016_janzing_chaves_scholkopf}: Suppose that $U_t \ket{\phi}$ had less quantum algorithmic information than $\ket{\phi}$. By definition there would be a circuit $\mathcal{C}_{U_t \phi}$, the concatenation of $\mathcal{C}_{U_t}$ and $\mathcal{C}_\phi$, with a simpler binary encoding than $\mathcal{C}_{\phi}$.\footnote{Time evolution for a large class of Hamiltonians can be well-approximated as a quantum circuit $\mathcal{C}_{U_t}$; see \citep[section 4.7]{2010_nielsen_chuang}.} That would entail one could obtain a shorter description of the initial circuit $\mathcal{C}_{\phi}$ by applying $\mathcal{C}_{U_t \phi}$ to the reference state $\ket{0}$ and appending $\mathcal{C}_{U^{-1}_t}$ to prepare $\ket{\phi}$: $\mathcal{C}_{U^{-1}_t}\mathcal{C}_{U_t \phi}\ket{0} = \ket{\phi}$. But by definition $\mathcal{C}_{\phi}$ is the shortest binary encoding of any circuit that will prepare $\ket{\phi}$ from $\ket{0}$, so this is impossible.%\footnote{It might seem blindingly obvious that $\mathcal{C}_{U^{-1}_t}\mathcal{C}_{U_t}\mathcal{C}_\phi$ would be more complex than $\mathcal{C}\phi$. However, it's possible that $\mathcal{C}_{U_t}$ could contain considerable information about $\mathcal{C}_\phi$, in which case $\mathcal{C}_{U^{-1}_t}\mathcal{C}_{U_t}\mathcal{C}_\phi \approx \mathcal{C}_\phi$ would be possible. That possibility is ruled out by $I(\ket{\phi} \, : \, U_t) = 0$.}

One may wonder how this is consistent with the measure-zero set of pure states whose Schmidt rank decreases for a given dynamical evolution U($t_i$). The answer is simple: they are excluded by imposing the Principle of Independent Mechanisms $I(\ket{\phi} \, : \, U_t) = 0$. Specifying a particular time interval $t_i$ of dynamical evolution tells one quite a lot about the candidate ``cause'' states for an entanglement-decreasing evolution. Indeed, it is only conditional on knowing U($t_i$) that one can even \textit{specify} such states. Even then they can only be picked out from the set of all pure states in $\mathcal{H}_N$ by including their dynamical behavior in their description. They are thus ruled out as candidate ``cause'' states because they fail to satisfy $I(\ket{\phi} \, : \, U_t) = 0$. 

From a physical perspective \textbf{No entanglement decrease} is quite reasonable. It would be baffling if a quantum system contained information about the dynamics, like the geometry and strength of the potential, \textit{prior} to having experienced them. It is obvious that the system will have acquired some information about the dynamics \textit{after} it is subjected to them. This informational asymmetry will be encoded in the quantum state -- in multiple ways, surely, but I have focused on the extent to which this information about the dynamics is captured by the degree of entanglement in the quantum state. This gives us a new criterion for causal inference: 

\begin{enumerate}
	\setcounter{enumi}{2}
\item[] \textit{Given two states $\ket{\alpha}$ and $\ket{\beta}$ related by unitary evolution, if $K^\varepsilon_Q(\ket{\alpha}) < K^\varepsilon_Q(\ket{\beta})$ then $\ket{\beta}$ was caused by time-evolving $\ket{\alpha}$.}
\end{enumerate}

\noindent This criterion represents an improvement in several respects on the criteria identified at the end of section 3. First, the introduction of a fidelity $\varepsilon$ means that it no longer depends on an unsatisfiable standard of precision, as noted above.

Second, the set of states for which it holds has finite measure in the set of all pure states in $\mathcal{H}_N$, as discussed above. The algorithmic information $K^\varepsilon_Q$ is defined on \textit{patches} of pure states of small, but finite, volume and so states of non-maximal algorithmic information will occupy finite volume in the set of all pure states in $\mathcal{H}_N$. A similar conclusion follows from recalling that in the set of all pure states, the relative frequency of pure states of maximal algorithmic complexity is $\approx 1 - 2^{2^N \log \varepsilon}$. 

It is illuminating to contrast this with the case of classical statistical mechanics; for example, \cite{2017_goldstein_et_al} estimate that for a system of $N \sim 10^{20}$ particles, the relative volume of phase space occupied by microstates in the equilibrium macrostate is $\approx 1 - e^{-10^{-15}N}$.\footnote{More precisely, this the volume occupied by the equilibrium macrostate for any energy shell in phase space containing microstates with energies in the range $E + \Delta E$.} The relative frequency of pure states of maximal algorithmic complexity thus represents a difference in degree from the classical statistical mechanical case -- possibly a dramatic degree, depending on the system size \textit{N} and chosen fidelity $\varepsilon$ -- but not a difference in kind. %In fact, while in principle $\varepsilon$ could be arbitrarily small, current state-of-the-art values for achievable fidelities on few-qubit systems in a laboratory setting are around $\mathcal{O}(10^{-4})$ \citep{2014_harty_et_al,2016_harty_et_al}, a value that entails both a non-trivial volume for each ``patch''  on the unit sphere of pure states in $\mathcal{H}_N$ and a non-trivial relative frequency of states of non-maximal algorithmic complexity in the set of all pure states.
%Give state-of-the-art values for fidelity using quantum gates.

Third, the criteria proposed in section 3 were too weak to identify a causal direction between two pure states each of less-than-full Schmidt rank, as noted in fn. \ref{NMSR}. The algorithmic information-theoretic criterion can do better: given two pure states of \textit{any} Schmidt measure, \textbf{No entanglement decrease} entails that the state with strictly lower algorithmic information is the cause state. Finally, the criteria in section 3 were restricted to bipartite systems, while this algorithmic information-theoretic criterion extends also to multipartite systems of qubits. Note, however, that it does \textit{not} straightforwardly extend to infinite-dimensional systems, while the criteria in section 3 applied to both finite and infinite-dimensional systems.

%SAY MORE ABOUT HOW PREPARATION COMPLEXITY MAKES THIS ARGUMENT WORK FOR NEIGHBORHOODS WHEN NORM-DISTANCE DID NOT MAKE IT WORK. WE'RE SHIFTING FROM THE ENTANGLEMENT PROPERTIES OF THE STATE ITSELF TO THE AMOUNT OF INFORMATION REQUIRED TO PREPARE $\ket{0}$ VERY CLOSE TO A STATE WITH THOSE ENTANGLEMENT PROPERTIES. I GUESS IT'S JUST THAT WE'RE DEFINING THE COMPLEXITY OF A NEIGHBORHOOD BY THE COMPLEXITY OF THE STATE AT ITS CENTER = THE STATE $\psi$ SUCH THAT EVERY STATE IN THE NEIGHBORHOOD IS WITHIN $\varepsilon$ of $\psi$ = THE STATE WE'RE TRYING TO APPROXIMATE. QUESTION: DOES THIS RESULT IN THE SAME STATE BEING ASSIGNED DIFFERENT COMPLEXITIES DEPENDING ON WHICH NEIGHBORHOOD WE'RE CONSIDERING IT TO BE A PART OF? IS THAT CONSISTENT WITH THE COMPLEXITY OF A STATE BEING AN INTRINSIC PROPERTY?

One may naturally wonder if \textbf{No entanglement decrease} is an artifact of the chosen definition of quantum algorithmic information. There is good reason to be optimistic on this score. Many proposals for extending algorithmic information to quantum states assign higher algorithmic information to entangled states (see \citep{2007_mora_briegel_kraus} for a review) so one can reasonably expect that some version of the conclusion here concerning $K^\varepsilon_Q$ could be reproduced for other proposed measures of quantum algorithmic information. Additionally, the measure here is defined using quantum gates -- the unitary operations composing the circuit $\mathcal{C}_\phi$ -- and one may be concerned that this limits its application to systems of qubits. That concern can be alleviated: there is no principled obstacle to extending the use of quantum gates beyond two-level systems to more interesting multi-level systems, or even to quantum systems with continuous degrees of freedom \cite{2005_braunstein_van_loock}, and thus to extending the notion of algorithmic information employed in this paper to those systems. %Indeed, quantum gates have recently been used to extend to quantum field theory the notion of \textit{circuit complexity}, which has much in common with the definition of quantum algorithmic complexity used here \citep{2017_jefferson_myers}.

There is a significant limitation to this result, however, which is a well-known drawback of any use of algorithmic information theory: the algorithmic complexity of an object is not computable. Making use of this causal inference criterion in any practical setting would thus require identifying contextually appropriate, computable substitutes for the algorithmic complexity. Certain computable substitutes have already been proposed in different contexts; for example, the ``information-geometric approach'' of \citep{2010_daniusis_janzing_mooij,2012_daniusis_janzing}. (The practical applicability of this method is itself limited by the fact that it requires deterministic relations between cause and effect variables.) It is also the case that the particular scenario here -- inferring the causal ordering of two known pure states -- is itself of limited practical application in the laboratory; after all, one doesn't typically find pure states lying around in the world (to borrow a phrase from a referee). As a philosophical result, it succeed in demonstrating that contrary to what Russell and modern-day Russellians say about causation in ``fundamental’’ physics, the time-symmetric nature of dynamical evolution does not make it impossible for that dynamical evolution to ground relations of cause and effect. However, as a procedure intended for practical use in the laboratory, this is only a small first step in applying the methods of Janzing, Sch\"olkopf, and collaborators to a quantum context.

That said, focusing solely on the strength of the causal inference criterion underwritten by \textbf{No entanglement decrease} may be too parochial. Arguably the primary interest of the criterion proposed here lies as much in the method used to reach it as its content. By attributing to quantum states a measure of algorithmic information and incorporating them into an empirically successful framework for causal inference, one can identify causal asymmetries between quantum states related by unitary time evolution. All that is required is a very weak assumption: the Principle of Independent Mechanisms. The extension of the methods of \citep{2008_janzing_scholkopf,2017_peters_janzing_scholkopf} to quantum theories introduces a novel set of mathematical and conceptual tools into the landscape for quantum causal inference. The fact that those methods confirm a conclusion that can also be reached on physical grounds provides a comforting verification of their reliability in this context, while their ability to extend that conclusion in select ways encourages their broader exploration.

\section{Conclusion}

In this paper I have argued that one can give a general and principled identification of a causal asymmetry between pure states of certain quantum systems related by unitary time evolution. The physical fact underlying this identification is simple: interaction between quantum systems entangles those systems but almost never disentangles them. Although there are many ways to make that physical fact precise, I have focused on one in particular for identifying causal asymmetry: the Schmidt rank (more generally, the Schmidt measure) of the two quantum states in question. I argued for the presence of this asymmetry on physical grounds in section 3, while in section 5 I did so by applying algorithmic information-theoretic concepts and methods from causal inference. In both cases, the Principle of Independent Mechanisms played a central role: the requirement that one ought to be able to specify the ``cause'' state without any information about the dynamics identified the causal asymmetry. Also in both cases, the asymmetry arose not in the laws governing dynamical evolution themselves, but between the states related by that dynamical evolution. In this respect, the asymmetry was identified by abandoning what \citep[section 11]{2020_woodward} calls the ``cause-in-laws'' attitude: the belief that if the laws of dynamical evolution themselves do not reflect an asymmetry, then there is no asymmetry to be found. Instead, my discussion here is a reflection of the more general fact that ``the directional features of causation are closely bound up with facts about the initial and boundary conditions of the systems we are analyzing and the way in which these are related to or interact with the [laws] governing those systems'' \citep[section 11]{2020_woodward}. That one can identify causal asymmetries in the dynamical evolution of quantum theories is indication that, \textit{pace} Russell and others, one can locate a meaningful notion of causation in microscopic physics after all.

I want to conclude by highlighting a couple of philosophical questions that the previous sections raise. The first is the relationship between causal asymmetry and temporal asymmetry. Couldn't the conclusions here about causal asymmetry be simply recast as conclusions about inferring temporal asymmetry? That is, couldn't this have been described as an exercies in determining which of $\ket{\alpha}$ or $\ket{\beta}$ was the \textit{initial} or \textit{earlier} state, rather than whether $\ket{\beta}$ was caused by time-evolving $\ket{\alpha}$ or vice versa? This is certainly how Janzing, Sch\"olkopf, and collaborators interpret much of their work; when applied to a time series or within the context of physical theory, for instance, they frequently describe their aim interchangeably as distinguishing between cause and effect or between \textit{past} and \textit{future}.%\footnote{This has occasionally met with amusing feedback: 

%\begin{quote}
%	We submitted the paper about deciding about the direction of time series -- can we tell whether a time series is forward or backwards? And some of the reviewers said this is a completely stupid question. Why would anyone be interested in whether a time series is forward or backwards? \citep[1:00:11]{2013_scholkopf_janzing}.
%\end{quote}
%}

I think that it is quite reasonable to accept that \textit{in the setup I have considered}, a causal asymmetry entails a temporal asymmetry. Indeed, one might think that when applied to states and dynamics in physical theories, some pre-theoretic understanding that causal asymmetries are almost always asymmetries \textit{in time} is what lends the Principle of Independent Mechanisms its plausibility. However, I ultimately do think that what is being inferred from these methods is, first and foremost, a causal relationship. One reason for thinking this is that the methods of Janzing et al. that I use in this paper have been shown to perform well at distinguishing cause from effect even for data sets that are not time series data (e.g. the relationship between altitude and temperature, or longitude and precipitation in \citep{2016_mooji_peters_janzing}.). That suggests that the structure being identified by these methods is causal structure; it just so happens that in the restricted setup being considered in this paper, that causal asymmetry coincides with the temporal asymmetry.

%Decoherence just tracks increasing entanglement between a system and its environment: the non-unitary that occurs when the environment is traced out is the most obvious formal manifestation of this asymmetry, but I have argued that the asymmetry arises at a more basic level from the generation of entanglement, but almost never disentanglement, by the unitary dynamics. 
Second, one might wonder how this relates to other recognized temporal asymmetries in quantum theories. The clearest relationship is with decoherence.\footnote{See \cite[chapter 9]{2012_wallace} for a discussion of decoherence and temporal asymmetry.} As is well known, in quantum systems of many particles the entanglement between any two subsystems rapidly decoheres as each subsystem interacts, and becomes increasingly entangled, with other particles in the system.\footnote{In language that may be more familiar: typically one splits the multiparticle system into two and calls the subsystem(s) of interest ``the system'' and the rest of the particles ``the environment''. The entanglement between the subsystems of interest is then decohered by their respective entanglements with the environment.} It essentially never happens that a multiparticle system that is initially, or evolves into, a highly entangled state is  later \textit{disentangled} through some series of interactions between the particles. This is a manifestation of the general physical fact at the foundation of this paper: interactions between quantum systems generically entangle those systems but almost never disentangle them. The non-unitarity of time evolution associated with decoherence that arises when one traces out all but the subystem(s) of interest is the sharpest formal sense in which decoherence is asymmetric in time. There is an asymmetry that arises prior to the tracing out, however: the increasing entanglement between subsystems.\footnote{As I mentioned in section 1, the relationship between entanglement and the direction of time has received extensive exploration.}

With this in mind, one could focus on the physical fact underlying my formal discussion of causal inference: the generic creation of entanglement between quantum systems by interactions. One could then view my discussion as merely one way to formalize that fact in a restricted context, in a way chosen specificially to be amenable to a particular set of formal methods of causal inference. In that case, one could see that physical fact as identifying an asymmetry that is conceptually prior both to the asymmetry I identifed using causal inference methods \textit{and} to the asymmetry introduced when, after two entangled subsystems are decohered through interactions with some ``environment'', that environment is traced out to produce a non-unitary evolution for the subsystems of interest. With that attitude, the causal asymmetry I have identified for bipartite (and multipartite) pure states is just the maximally microscopic manifestation of the very same physical conditions that produce the asymmetry manifest in decoherence.

Finally, I have spoken consistently of \textit{inferring} or \textit{identifying} causal asymmetry. As this epistemic language suggests, I have not primarily been concerned with providing an analysis of what causation or causal asymmetry \textit{is}, in any strong metaphysical sense, but rather with certain circumstances in quantum theories under which one can identify it. That said, empirically accessible manifestations of an object in, or property of, the natural world -- and thus successful methods for detecting that object or property -- are not independent of the way that object, or that property, \textit{actually is}. By focusing on how one can learn about causation under particular circumstances, one can hope to learn something about the worldly properties that provide the physical basis for, or ground, our reasoning about causal relations in various contexts. I am particularly influenced on this score by \citep{2020_woodward}:\footnote{See also \citep[p. 914]{2009_eberhardt}: 
	
	\begin{quote}``$\ldots$ metaphysical accounts have provided essentially no guidance for methods of discovery because it remains unclear how they could be operationalized into discovery procedures that do not depend on the availability of causal knowledge in the first place. Epistemological headway was made by a completely different strategy that largely ignored metaphysical considerations.''\end{quote} 
	
	\noindent I would only emphasize that I think that once one has made this epistemological headway, it can be valuable to use it as a basis for circling back to address some of the metaphysical considerations that one initially set aside to make epistemic progress.}

\begin{quote}
	This is the project, alluded to earlier, of elucidating the worldly infrastructure that underlies and grounds assessments of causal direction. I see this project as connecting epistemological concerns having to do with how we find out about causal direction with the ``what is out there'' concerns of metaphysicians, although my answer to the what is out there question does not involve any kind of elaborate metaphysics. My general picture is that causal thinking ``works'' to the extent that it does because it picks up on or is supported by certain generic features of our world $\ldots$ \citep[p. 5]{2020_woodward}.
\end{quote}

The causal inference methods of Janzing, Sch\"olkopf, and collaborators, or of \citep{2009_pearl} or \citep{2000_spirtes_glymour_scheines}, represent relatively unified \textit{methodological} strategies for identifying causal relations in the world without presuming any unified \textit{metaphysical} ground of those relations. These methodological frameworks allow that the physical basis, or metaphysical, ground of any particular causal relationship identified by those methods will likely vary with the worldly situations to which those methods are applied. Aside from the satisfaction of certain statistical or algorithmic information-theoretic properties, there may be little in their respective worldly grounds to link them together \textit{as causal relationships}; the relationships may be metaphysically necessary or contingent, may be fundamental or emergent, and so on. The ability of these methodological frameworks to act as reliable diagnostics for such a diversely instantiated class of causal relations is part of their immense epistemic value. In the case I have considered, the worldly structure that grounds the identifiation of a causal asymmetry is a differential degree of entanglement in two states related by unitary time evolution. This asymmetry, in turn, is grounded in the physical fact that interactions between subsystems in quantum theories generically produce entanglement. The fact that one can identify this causal asymmetry using methods of causal inference reflects not merely a fact about causal epistemology, but a fact about the nature of the physical world.

\section*{Acknowledgments}

I would like to thank John Dougherty, Casey McCoy, Michael Miller, Siddarth Muthukrishnan, and an audience at Oxford University for helpful feedback. I would like to thank David Albert, Naftali Weinberg, Jim Woodward, and two very helpful referees for valuable and constructive feedback on a previous draft. I am especially indebted to Jim for conversations that sparked the idea for the paper and for subsequent encouragement and helpful discussion.

\bibliographystyle{plainnat}
\bibliography{causal_asymmetry}

\begin{thebibliography}{108}
\providecommand{\natexlab}[1]{#1}
\providecommand{\url}[1]{\texttt{#1}}
\expandafter\ifx\csname urlstyle\endcsname\relax
  \providecommand{\doi}[1]{doi: #1}\else
  \providecommand{\doi}{doi: \begingroup \urlstyle{rm}\Url}\fi

\bibitem[Albert(2000)]{2000_albert}
David~Z Albert.
\newblock \emph{Time and Chance}.
\newblock Harvard University Press, 2000.

\bibitem[Allen et~al.(2017)Allen, Barrett, Horsman, Lee, and
  Spekkens]{2017_barrett_spekkens_et_al}
John-Mark~A Allen, Jonathan Barrett, Dominic~C Horsman, Ciar{\'a}n~M Lee, and
  Robert~W Spekkens.
\newblock Quantum common causes and quantum causal models.
\newblock \emph{Physical Review X}, 7\penalty0 (3):\penalty0 031021, 2017.

\bibitem[Allori(2019)]{2019_allori}
Valia Allori.
\newblock Quantum mechanics, time and ontology.
\newblock \emph{Studies in History and Philosophy of Science Part B: Studies in
  History and Philosophy of Modern Physics}, 66:\penalty0 145--154, 2019.

\bibitem[Amico et~al.(2008)Amico, Fazio, Osterloh, and
  Vedral]{2008_amico_et_al}
Luigi Amico, Rosario Fazio, Andreas Osterloh, and Vlatko Vedral.
\newblock Entanglement in many-body systems.
\newblock \emph{Reviews of Modern Physics}, 80\penalty0 (2):\penalty0 517,
  2008.

\bibitem[Barrett et~al.(2019)Barrett, Lorenz, and
  Oreshkov]{2019_barrett_lorenz_oreshkov}
Jonathan Barrett, Robin Lorenz, and Ognyan Oreshkov.
\newblock Quantum causal models.
\newblock \emph{arXiv preprint arXiv:1906.10726}, 2019.

\bibitem[Berthiaume et~al.(2001)Berthiaume, Van~Dam, and
  Laplante]{2001_berthiaume_van_dam_laplante}
Andr{\'e} Berthiaume, Wim Van~Dam, and Sophie Laplante.
\newblock Quantum kolmogorov complexity.
\newblock \emph{Journal of Computer and System Sciences}, 63\penalty0
  (2):\penalty0 201--221, 2001.

\bibitem[Binney and Skinner(2013)]{2013_binney_skinner}
James Binney and David Skinner.
\newblock \emph{The physics of quantum mechanics}.
\newblock Oxford University Press, 2013.

\bibitem[Braunstein and Van~Loock(2005)]{2005_braunstein_van_loock}
Samuel~L Braunstein and Peter Van~Loock.
\newblock Quantum information with continuous variables.
\newblock \emph{Reviews of Modern Physics}, 77\penalty0 (2):\penalty0 513,
  2005.

\bibitem[Brock(2005)]{2005_brock}
Kerry~G Brock.
\newblock How rare are singular matrices?
\newblock \emph{The Mathematical Gazette}, 89\penalty0 (516):\penalty0
  378--384, 2005.

\bibitem[Bru{\ss}(2002)]{2001_bruss}
Dagmar Bru{\ss}.
\newblock Characterizing entanglement.
\newblock \emph{Journal of Mathematical Physics}, 43\penalty0 (9):\penalty0
  4237--4251, 2002.

\bibitem[Calabrese and Cardy(2006)]{2006_calabrese_cardy}
Pasquale Calabrese and John Cardy.
\newblock Time dependence of correlation functions following a quantum quench.
\newblock \emph{Physical Review Letters}, 96\penalty0 (13):\penalty0 136801,
  2006.

\bibitem[Callender(2000)]{2000_callender}
Craig Callender.
\newblock Is time `handed' in a quantum world?
\newblock \emph{Proceedings of the Aristotelian Society}, 100\penalty0
  (1):\penalty0 247--269, 2000.

\bibitem[Callender(2020)]{2020_callender}
Craig Callender.
\newblock Quantum mechanics: Keeping it real?
\newblock \emph{philsci-archive preprint:17701}, 2020.

\bibitem[Carroll and Singh(2020)]{2020_carroll_singh}
Sean~M Carroll and Ashmeet Singh.
\newblock Quantum mereology: Factorizing hilbert space into subsystems with
  quasi-classical dynamics.
\newblock \emph{arXiv preprint:2005.12938}, 2020.

\bibitem[Cervera-Lierta et~al.(2017)Cervera-Lierta, Latorre, Rojo, and
  Rottoli]{2017_cervera_lieta_et_al}
Alba Cervera-Lierta, Jos{\'e}~I Latorre, Juan Rojo, and Luca Rottoli.
\newblock Maximal entanglement in high energy physics.
\newblock \emph{SciPost Physics}, 3, 2017.

\bibitem[Chaves et~al.(2014)Chaves, Luft, and Gross]{2014_chaves_luft_gross}
Rafael Chaves, Lukas Luft, and David Gross.
\newblock Causal structures from entropic information: geometry and novel
  scenarios.
\newblock \emph{New Journal of Physics}, 16\penalty0 (4):\penalty0 043001,
  2014.

\bibitem[Chaves et~al.(2015)Chaves, Majenz, and
  Gross]{2015_chaves_majenz_gross}
Rafael Chaves, Christian Majenz, and David Gross.
\newblock Information--theoretic implications of quantum causal structures.
\newblock \emph{Nature communications}, 6\penalty0 (1):\penalty0 1--8, 2015.

\bibitem[Chiribella et~al.(2009)Chiribella, D’Ariano, and
  Perinotti]{2009_chiribella_et_al}
Giulio Chiribella, Giacomo~Mauro D’Ariano, and Paolo Perinotti.
\newblock Theoretical framework for quantum networks.
\newblock \emph{Physical Review A}, 80\penalty0 (2):\penalty0 022339, 2009.

\bibitem[Clifton and Halvorson(1999)]{1999_clifton_halvorson}
Rob Clifton and Hans Halvorson.
\newblock Bipartite-mixed-states of infinite-dimensional systems are
  generically nonseparable.
\newblock \emph{Physical Review A}, 61\penalty0 (1):\penalty0 012108, 1999.

\bibitem[Costa and Shrapnel(2016)]{2016_costa_shrapnel}
Fabio Costa and Sally Shrapnel.
\newblock Quantum causal modelling.
\newblock \emph{New Journal of Physics}, 18\penalty0 (6):\penalty0 063032,
  2016.

\bibitem[Cotler et~al.(2019)Cotler, Penington, and
  Ranard]{2019_cotler_penington_renard}
Jordan~S Cotler, Geoffrey~R Penington, and Daniel~H Ranard.
\newblock Locality from the spectrum.
\newblock \emph{Communications in Mathematical Physics}, 368\penalty0
  (3):\penalty0 1267--1296, 2019.

\bibitem[Daniu{\v s}is et~al.(2010)Daniu{\v s}is, Janzing, Mooij, Zscheischler,
  Steudel, Zhang, and Sch{\"o}lkopf]{2010_daniusis_janzing_mooij}
Povilas Daniu{\v s}is, Dominik Janzing, Joris~M. Mooij, Jakob Zscheischler,
  Bastian Steudel, Kun Zhang, and Bernhard Sch{\"o}lkopf.
\newblock Inferring deterministic causal relations.
\newblock In \emph{Proceedings of the 26th Annual Conference on {U}ncertainty
  in {A}rtificial {I}ntelligence ({UAI}-10)}, 2010.

\bibitem[Di~Biagio et~al.(2021)Di~Biagio, Don{\`a}, and
  Rovelli]{2021_rovelli_et_al}
Andrea Di~Biagio, Pietro Don{\`a}, and Carlo Rovelli.
\newblock The arrow of time in operational formulations of quantum theory.
\newblock \emph{Quantum}, 5:\penalty0 520, 2021.

\bibitem[Donoghue and Menezes(2019)]{2019_donoghue_menezes}
John Donoghue and Gabriel Menezes.
\newblock Arrow of causality and quantum gravity.
\newblock \emph{Physical Review Letters}, 123\penalty0 (17):\penalty0 171601,
  2019.

\bibitem[Donoghue and Menezes(2020)]{2020_donoghue_menezes}
John Donoghue and Gabriel Menezes.
\newblock Quantum causality determines the arrow of time.
\newblock \emph{arXiv preprint:2003.09047}, 2020.

\bibitem[Earman(2002)]{2002_earman}
John Earman.
\newblock What time reversal invariance is and why it matters.
\newblock \emph{International Studies in the Philosphy of Science}, 16\penalty0
  (3):\penalty0 245--264, 2002.

\bibitem[Earman(2015)]{2015_earman}
John Earman.
\newblock Some puzzles and unresolved issues about quantum entanglement.
\newblock \emph{Erkenntnis}, 80\penalty0 (2):\penalty0 303--337, 2015.

\bibitem[Eberhardt(2009)]{2009_eberhardt}
Frederick Eberhardt.
\newblock Introduction to the epistemology of causation.
\newblock \emph{Philosophy Compass}, 4\penalty0 (6):\penalty0 913--925, 2009.

\bibitem[Eisert and Briegel(2001)]{2001_eisert_briegel}
Jens Eisert and Hans~J Briegel.
\newblock Schmidt measure as a tool for quantifying multiparticle entanglement.
\newblock \emph{Physical Review A}, 64\penalty0 (2):\penalty0 022306, 2001.

\bibitem[Eisert and Osborne(2006)]{2006_eisert_osborne}
Jens Eisert and Tobias~J Osborne.
\newblock General entanglement scaling laws from time evolution.
\newblock \emph{Physical Review Letters}, 97\penalty0 (15):\penalty0 150404,
  2006.

\bibitem[Englert et~al.(1988)Englert, Schwinger, and
  Scully]{1988_schwinger_englert_scully}
Berthold-Georg Englert, Julian Schwinger, and Marlan~O Scully.
\newblock Is spin coherence like {H}umpty-{D}umpty? {I}: {S}implified
  treatment.
\newblock \emph{Foundations of Physics}, 18\penalty0 (10):\penalty0 1045--1056,
  1988.

\bibitem[Farr(2020)]{2020_farr}
Matt Farr.
\newblock Causation and time reversal.
\newblock \emph{The British Journal for the Philosophy of Science}, 71\penalty0
  (1):\penalty0 177--204, 2020.

\bibitem[Farr and Reutlinger(2013)]{2013_farr_reutlinger}
Matt Farr and Alexander Reutlinger.
\newblock A relic of a bygone age? causation, time symmetry and the
  directionality argument.
\newblock \emph{Erkenntnis}, 78\penalty0 (2):\penalty0 215--235, 2013.

\bibitem[Fernandes(2017)]{2017_fernandes}
Alison Fernandes.
\newblock A deliberative approach to causation.
\newblock \emph{Philosophy and Phenomenological Research}, 95\penalty0
  (3):\penalty0 686--708, 2017.

\bibitem[Field(2003)]{2003_field}
Hartry Field.
\newblock Causation in a physical world.
\newblock In Michael~J. Loux and Dean Zimmerman, editors, \emph{Oxford Handbook
  of Metaphysics}, pages 435--60. Oxford University Press, Oxford, 2003.

\bibitem[Fine(1982)]{1982_fine}
Arthur Fine.
\newblock Joint distributions, quantum correlations, and commuting observables.
\newblock \emph{Journal of Mathematical Physics}, 23\penalty0 (7):\penalty0
  1306--1310, 1982.

\bibitem[Goldstein et~al.(2013)Goldstein, Hara, and
  Tasaki]{2013_goldstein_hara_tasaki}
Sheldon Goldstein, Takashi Hara, and Hal Tasaki.
\newblock Time scales in the approach to equilibrium of macroscopic quantum
  systems.
\newblock \emph{Physical review letters}, 111\penalty0 (14):\penalty0 140401,
  2013.

\bibitem[Goldstein et~al.(2015{\natexlab{a}})Goldstein, Hara, and
  Tasaki]{2015_goldstein_hara_tasaki}
Sheldon Goldstein, Takashi Hara, and Hal Tasaki.
\newblock Extremely quick thermalization in a macroscopic quantum system for a
  typical nonequilibrium subspace.
\newblock \emph{New Journal of Physics}, 17\penalty0 (4):\penalty0 045002,
  2015{\natexlab{a}}.

\bibitem[Goldstein et~al.(2015{\natexlab{b}})Goldstein, Huse, Lebowitz, and
  Tumulka]{2015_goldstein_et_al}
Sheldon Goldstein, David~A Huse, Joel~L Lebowitz, and Roderich Tumulka.
\newblock Thermal equilibrium of a macroscopic quantum system in a pure state.
\newblock \emph{Physical Review Letters}, 115\penalty0 (10):\penalty0 100402,
  2015{\natexlab{b}}.

\bibitem[Goldstein et~al.(2017)Goldstein, Huse, Lebowitz, and
  Tumulka]{2017_goldstein_et_al}
Sheldon Goldstein, David~A Huse, Joel~L Lebowitz, and Roderich Tumulka.
\newblock Macroscopic and microscopic thermal equilibrium.
\newblock \emph{Annalen der Physik}, 529\penalty0 (7):\penalty0 1600301, 2017.

\bibitem[Grunwald and Vit{\'a}nyi(2010)]{2010_grunwald_vitanyi}
Peter Grunwald and Paul Vit{\'a}nyi.
\newblock Shannon information and kolmogorov complexity.
\newblock \emph{arXiv preprint:cs/0410002}, 2010.

\bibitem[Hall(2013)]{2013_hall}
Brian~C Hall.
\newblock \emph{Quantum theory for mathematicians}.
\newblock Springer, 2013.

\bibitem[Hardy(2021)]{2021_hardy}
Lucien Hardy.
\newblock Time symmetry in operational theories.
\newblock \emph{arXiv preprint arXiv:2104.00071}, 2021.

\bibitem[Hein et~al.(2004)Hein, Eisert, and Briegel]{2004_hein_eisert_briegel}
Marc Hein, Jens Eisert, and Hans~J Briegel.
\newblock Multiparty entanglement in graph states.
\newblock \emph{Physical Review A}, 69\penalty0 (6):\penalty0 062311, 2004.

\bibitem[Ismael(2016)]{2016_ismael}
Jenann Ismael.
\newblock \emph{How physics makes us free}.
\newblock Oxford University Press, 2016.

\bibitem[Janzing(2019)]{2006_janzing}
Dominik Janzing.
\newblock \emph{Computer science approach to quantum control}.
\newblock KIT Scientific Publishing, 2019.

\bibitem[Janzing and Sch{\"o}lkopf(2010)]{2008_janzing_scholkopf}
Dominik Janzing and Bernhard Sch{\"o}lkopf.
\newblock Causal inference using the algorithmic markov condition.
\newblock \emph{IEEE Transactions on Information Theory}, 56\penalty0
  (10):\penalty0 5168--5194, 2010.

\bibitem[Janzing and Wocjan(2018)]{2018_janzing_wocjan}
Dominik Janzing and Pawe{\l} Wocjan.
\newblock Does universal controllability of physical systems prohibit
  thermodynamic cycles?
\newblock \emph{Open Systems \& Information Dynamics}, 25\penalty0
  (03):\penalty0 1850016, 2018.

\bibitem[Janzing et~al.(2002)Janzing, Wocjan, and
  Beth]{2002_janzing_wocjan_beth}
Dominik Janzing, Pawel Wocjan, and Thomas Beth.
\newblock Complexity of decoupling and time reversal for n spins with pair
  interactions: Arrow of time in quantum control.
\newblock \emph{Physical Review A}, 66\penalty0 (4):\penalty0 042311, 2002.

\bibitem[Janzing et~al.(2012)Janzing, Mooij, Zhang, Lemeire, Zscheischler,
  Daniu{\v{s}}is, Steudel, and Sch{\"o}lkopf]{2012_daniusis_janzing}
Dominik Janzing, Joris Mooij, Kun Zhang, Jan Lemeire, Jakob Zscheischler,
  Povilas Daniu{\v{s}}is, Bastian Steudel, and Bernhard Sch{\"o}lkopf.
\newblock Information-geometric approach to inferring causal directions.
\newblock \emph{Artificial Intelligence}, 182:\penalty0 1--31, 2012.

\bibitem[Janzing et~al.(2016)Janzing, Chaves, and
  Sch{\"o}lkopf]{2016_janzing_chaves_scholkopf}
Dominik Janzing, Rafael Chaves, and Bernhard Sch{\"o}lkopf.
\newblock Algorithmic independence of initial condition and dynamical law in
  thermodynamics and causal inference.
\newblock \emph{New Journal of Physics}, 18\penalty0 (9):\penalty0 093052,
  2016.

\bibitem[Jennings and Rudolph(2010)]{2010_jennings_rudolph}
David Jennings and Terry Rudolph.
\newblock Entanglement and the thermodynamic arrow of time.
\newblock \emph{Physical Review E}, 81\penalty0 (6):\penalty0 061130, 2010.

\bibitem[Kharzeev and Levin(2017)]{2017_kharzeev_levin}
Dmitri~E Kharzeev and Eugene~M Levin.
\newblock Deep inelastic scattering as a probe of entanglement.
\newblock \emph{Physical Review D}, 95\penalty0 (11):\penalty0 114008, 2017.

\bibitem[Leifer and Spekkens(2013)]{2013_leifer_spekkens}
Matthew~S Leifer and Robert~W Spekkens.
\newblock Towards a formulation of quantum theory as a causally neutral theory
  of bayesian inference.
\newblock \emph{Physical Review A}, 88\penalty0 (5):\penalty0 052130, 2013.

\bibitem[Li and Vit{\'a}nyi(2019)]{2019_li_vitanyi}
Ming Li and Paul Vit{\'a}nyi.
\newblock \emph{An introduction to Kolmogorov complexity and its applications:
  Fourth edition}.
\newblock Springer, 2019.

\bibitem[Linden et~al.(2009)Linden, Popescu, Short, and
  Winter]{2009_linden_popescu_short_winter}
Noah Linden, Sandu Popescu, Anthony~J Short, and Andreas Winter.
\newblock Quantum mechanical evolution towards thermal equilibrium.
\newblock \emph{Physical Review E}, 79\penalty0 (6):\penalty0 061103, 2009.

\bibitem[Loewer(2012)]{2012_loewer}
Barry Loewer.
\newblock Two accounts of laws and time.
\newblock \emph{Philosophical Studies}, 160\penalty0 (1):\penalty0 115--137,
  2012.

\bibitem[Malabarba et~al.(2014)Malabarba, Garc{\'\i}a-Pintos, Linden, Farrelly,
  and Short]{2014_malabarba_et_al}
Artur~SL Malabarba, Luis~Pedro Garc{\'\i}a-Pintos, Noah Linden, Terence~C
  Farrelly, and Anthony~J Short.
\newblock Quantum systems equilibrate rapidly for most observables.
\newblock \emph{Physical Review E}, 90\penalty0 (1):\penalty0 012121, 2014.

\bibitem[Maudlin(2007)]{2007_maudlin}
Tim Maudlin.
\newblock \emph{The Metaphysics within Physics}.
\newblock Oxford University Press, 2007.

\bibitem[Meek(1995)]{1995_meek}
Chris Meek.
\newblock Strong-completeness and faithfulness in belief networks.
\newblock In \emph{Proceedings of the Eleventh Conference on Uncertainty in
  Artificial Intelligence}, 1995.

\bibitem[Mishima et~al.(2004)Mishima, Hayashi, and
  Lin]{2004_mishima_hayashi_lin}
K~Mishima, M~Hayashi, and SH~Lin.
\newblock Entanglement in scattering processes.
\newblock \emph{Physics Letters A}, 333\penalty0 (5-6):\penalty0 371--377,
  2004.

\bibitem[Mooij et~al.(2016)Mooij, Peters, Janzing, Zscheischler, and
  Sch{\"o}lkopf]{2016_mooji_peters_janzing}
Joris~M Mooij, Jonas Peters, Dominik Janzing, Jakob Zscheischler, and Bernhard
  Sch{\"o}lkopf.
\newblock Distinguishing cause from effect using observational data: methods
  and benchmarks.
\newblock \emph{The Journal of Machine Learning Research}, 17\penalty0
  (1):\penalty0 1103--1204, 2016.

\bibitem[Mora and Briegel(2005)]{2005_mora_briegel}
Caterina~E Mora and Hans~J Briegel.
\newblock Algorithmic complexity and entanglement of quantum states.
\newblock \emph{Physical Review Letters}, 95\penalty0 (20):\penalty0 200503,
  2005.

\bibitem[Mora and Briegel(2006)]{2006_mora_briegel}
Caterina~E Mora and Hans~J Briegel.
\newblock Algorithmic complexity of quantum states.
\newblock \emph{International Journal of Quantum Information}, 4\penalty0
  (04):\penalty0 715--737, 2006.

\bibitem[Mora et~al.(2007)Mora, Briegel, and Kraus]{2007_mora_briegel_kraus}
Caterina~E Mora, Hans~J Briegel, and Barbara Kraus.
\newblock Quantum kolmogorov complexity and its applications.
\newblock \emph{International Journal of Quantum Information}, 5\penalty0
  (05):\penalty0 729--750, 2007.

\bibitem[Mueller(2007)]{2007_mueller}
Markus Mueller.
\newblock Quantum kolmogorov complexity and the quantum turing machine.
\newblock \emph{arXiv preprint:0712.4377}, 2007.

\bibitem[Myrvold(2020)]{2020_myrvold}
Wayne~C Myrvold.
\newblock Explaining thermodynamics: What remains to be done?
\newblock In Valia Allori, editor, \emph{Statistical Mechanics And Scientific
  Explanation: Determinism, Indeterminism And Laws Of Nature}, pages 113--143.
  World Scientific, 2020.

\bibitem[Nielsen and Chuang(2010)]{2010_nielsen_chuang}
Michael Nielsen and Isaac Chuang.
\newblock \emph{Quantum Computation and Quantum Information: 10th Anniversary
  Edition}.
\newblock Cambridge University Press, 2010.

\bibitem[Oreshkov et~al.(2012)Oreshkov, Costa, and
  Brukner]{2012_oreshkov_et_al}
Ognyan Oreshkov, Fabio Costa, and {\v{C}}aslav Brukner.
\newblock Quantum correlations with no causal order.
\newblock \emph{Nature communications}, 3\penalty0 (1):\penalty0 1--8, 2012.

\bibitem[Pearl(2009)]{2009_pearl}
Judea Pearl.
\newblock \emph{Causality: 2nd Edition}.
\newblock Cambridge University Press, 2009.

\bibitem[Peschanski and Seki(2016)]{2016_peschanski_seki}
Robi Peschanski and Shigenori Seki.
\newblock Entanglement entropy of scattering particles.
\newblock \emph{Physics Letters B}, 758:\penalty0 89--92, 2016.

\bibitem[Peters et~al.(2017)Peters, Janzing, and
  Sch{\"o}lkopf]{2017_peters_janzing_scholkopf}
Jonas Peters, Dominik Janzing, and Bernhard Sch{\"o}lkopf.
\newblock \emph{Elements of Causal Inference}.
\newblock The MIT Press, 2017.

\bibitem[Popescu et~al.(2006)Popescu, Short, and
  Winter]{2006_popescu_short_winter}
Sandu Popescu, Anthony~J Short, and Andreas Winter.
\newblock Entanglement and the foundations of statistical mechanics.
\newblock \emph{Nature Physics}, 2\penalty0 (11):\penalty0 754--758, 2006.

\bibitem[Preskill(1998)]{1998_preskill}
John Preskill.
\newblock Lecture notes for physics 229: Quantum information and computation.
\newblock \emph{Available at http://theory.caltech.edu/~preskill/ph229/}, 1998.

\bibitem[Price(2007)]{2007_price}
Huw Price.
\newblock Causal perspectivalism.
\newblock In Huw Price and Richard Corry, editors, \emph{Causation, physics,
  and the constitution of reality: Russell's republic revisited}. Oxford
  University Press, Oxford, 2007.

\bibitem[Reimann(2008)]{2008_reimann}
Peter Reimann.
\newblock Foundation of statistical mechanics under experimentally realistic
  conditions.
\newblock \emph{Physical review letters}, 101\penalty0 (19):\penalty0 190403,
  2008.

\bibitem[Ried et~al.(2015)Ried, Agnew, Vermeyden, Janzing, Spekkens, and
  Resch]{2015_ried_janzing_spekkens_et_al}
Katja Ried, Megan Agnew, Lydia Vermeyden, Dominik Janzing, Robert~W Spekkens,
  and Kevin~J Resch.
\newblock A quantum advantage for inferring causal structure.
\newblock \emph{Nature Physics}, 11\penalty0 (5):\penalty0 414--420, 2015.

\bibitem[Roberts(2017)]{2017_roberts}
Bryan~W Roberts.
\newblock Three myths about time reversal in quantum theory.
\newblock \emph{Philosophy of Science}, 84\penalty0 (2):\penalty0 315--334,
  2017.

\bibitem[Roberts(2019)]{2019_roberts}
Bryan~W Roberts.
\newblock Time reversal.
\newblock \emph{philsci-archive preprint:15033}, 2019.

\bibitem[Rubino et~al.(2017)Rubino, Rozema, Feix, Ara{\'u}jo, Zeuner, Procopio,
  Brukner, and Walther]{2017_brukner_et_al}
Giulia Rubino, Lee~A Rozema, Adrien Feix, Mateus Ara{\'u}jo, Jonas~M Zeuner,
  Lorenzo~M Procopio, {\v{C}}aslav Brukner, and Philip Walther.
\newblock Experimental verification of an indefinite causal order.
\newblock \emph{Science advances}, 3\penalty0 (3):\penalty0 e1602589, 2017.

\bibitem[Rudin(1987)]{1987_rudin}
Walter Rudin.
\newblock \emph{Real and Complex Analysis}.
\newblock McGraw-Hill, 1987.

\bibitem[Russell(1912)]{1912_russell}
Bertrand Russell.
\newblock On the notion of cause.
\newblock \emph{Proceedings of the Aristotelian society}, 13:\penalty0 1--26,
  1912.

\bibitem[Sachs(1987)]{1987_sachs}
Robert~G Sachs.
\newblock \emph{The physics of time reversal}.
\newblock University of Chicago Press, 1987.

\bibitem[Sakurai and Napolitano(2011)]{2011_sakurai}
J.J. Sakurai and Jim Napolitano.
\newblock \emph{Quantum Mechanics: Second Edition}.
\newblock Addison-Wesley, 2011.

\bibitem[Sanpera et~al.(2001)Sanpera, Bru{\ss}, and
  Lewenstein]{2001_sanpera_bruss_lewenstein}
Anna Sanpera, Dagmar Bru{\ss}, and Maciej Lewenstein.
\newblock Schmidt-number witnesses and bound entanglement.
\newblock \emph{Physical Review A}, 63\penalty0 (5):\penalty0 050301, 2001.

\bibitem[Schmid et~al.(2020)Schmid, Selby, and
  Spekkens]{2020_schmid_selby_spekkens}
David Schmid, John~H Selby, and Robert~W Spekkens.
\newblock Unscrambling the omelette of causation and inference: The framework
  of causal-inferential theories.
\newblock \emph{arXiv preprint arXiv:2009.03297}, 2020.

\bibitem[Schroeder(2017)]{2017_schroeder}
Daniel~V Schroeder.
\newblock Entanglement isn't just for spin.
\newblock \emph{American Journal of Physics}, 85\penalty0 (11):\penalty0
  812--820, 2017.

\bibitem[Short and Farrelly(2012)]{2012_short_farrelly}
Anthony~J Short and Terence~C Farrelly.
\newblock Quantum equilibration in finite time.
\newblock \emph{New Journal of Physics}, 14\penalty0 (1):\penalty0 013063,
  2012.

\bibitem[Sklar(1993)]{1993_sklar}
Lawrence Sklar.
\newblock \emph{Physics and Chance}.
\newblock Cambridge University Press, 1993.

\bibitem[Sperling and Vogel(2011)]{2011_sperling_vogel}
J~Sperling and W~Vogel.
\newblock The schmidt number as a universal entanglement measure.
\newblock \emph{Physica Scripta}, 83\penalty0 (4):\penalty0 045002, 2011.

\bibitem[Spirtes et~al.(2000)Spirtes, Glymour, Scheines, and
  Heckerman]{2000_spirtes_glymour_scheines}
Peter Spirtes, Clark~N Glymour, Richard Scheines, and David Heckerman.
\newblock \emph{Causation, prediction, and search}.
\newblock The MIT Press, 2000.

\bibitem[Struyve(2020)]{2020_struyve}
Ward Struyve.
\newblock Time-reversal invariance and ontology.
\newblock \emph{philsci-archive preprint:17682}, 2020.

\bibitem[Terhal and Horodecki(2000)]{2000_terhal_horodecki}
Barbara~M Terhal and Pawe{\l} Horodecki.
\newblock Schmidt number for density matrices.
\newblock \emph{Physical Review A}, 61\penalty0 (4):\penalty0 040301, 2000.

\bibitem[Thompson et~al.(2018)Thompson, Garner, Mahoney, Crutchfield, Vedral,
  and Gu]{2018_thompson_et_al}
Jayne Thompson, Andrew~JP Garner, John~R Mahoney, James~P Crutchfield, Vlatko
  Vedral, and Mile Gu.
\newblock Causal asymmetry in a quantum world.
\newblock \emph{Physical Review X}, 8\penalty0 (3):\penalty0 031013, 2018.

\bibitem[Timpson(2013)]{2013_timpson}
Christopher~G Timpson.
\newblock \emph{Quantum information theory and the foundations of quantum
  mechanics}.
\newblock OUP Oxford, 2013.

\bibitem[Van~den Nest(2013)]{2013_den_nest}
Maarten Van~den Nest.
\newblock Universal quantum computation with little entanglement.
\newblock \emph{Physical Review Letters}, 110\penalty0 (6):\penalty0 060504,
  2013.

\bibitem[Vit{\'a}nyi(2001)]{2001_vitanyi}
Paul~MB Vit{\'a}nyi.
\newblock Quantum kolmogorov complexity based on classical descriptions.
\newblock \emph{IEEE Transactions on Information Theory}, 47\penalty0
  (6):\penalty0 2464--2479, 2001.

\bibitem[Wallace(2012)]{2012_wallace}
David Wallace.
\newblock \emph{The Emergent Multiverse: Quantum theory according to the
  Everett interpretation}.
\newblock Oxford University Press, 2012.

\bibitem[Wallace(2015)]{2015_wallace}
David Wallace.
\newblock Recurrence theorems: a unified account.
\newblock \emph{Journal of Mathematical Physics}, 56\penalty0 (2):\penalty0
  022105, 2015.

\bibitem[Weinberger(2018)]{2018_weinberger}
Naftali Weinberger.
\newblock Faithfulness, coordination and causal coincidences.
\newblock \emph{Erkenntnis}, 83\penalty0 (2):\penalty0 113--133, 2018.

\bibitem[Wood and Spekkens(2015)]{2015_wood_spekkens}
Christopher~J Wood and Robert~W Spekkens.
\newblock The lesson of causal discovery algorithms for quantum correlations:
  Causal explanations of bell-inequality violations require fine-tuning.
\newblock \emph{New Journal of Physics}, 17\penalty0 (3):\penalty0 033002,
  2015.

\bibitem[Woodward(2005)]{2005_woodward}
James Woodward.
\newblock \emph{Making things happen: A theory of causal explanation}.
\newblock Oxford university press, 2005.

\bibitem[Woodward(2014)]{2014_woodward}
James Woodward.
\newblock A functional account of causation; or, a defense of the legitimacy of
  causal thinking by reference to the only standard that matters—usefulness
  (as opposed to metaphysics or agreement with intuitive judgment).
\newblock \emph{Philosophy of Science}, 81\penalty0 (5):\penalty0 691--713,
  2014.

\bibitem[Woodward(2020)]{2020_woodward}
James Woodward.
\newblock Flagpoles anyone? causal and explanatory asymmetries.
\newblock \emph{philsci-archive preprint:17419}, 2020.

\bibitem[Woodward(2007)]{2007_woodward}
Jim Woodward.
\newblock Causation with a human face.
\newblock In Huw Price and Richard Corry, editors, \emph{Causation, Physics,
  and the Constitution of Reality: Russell's Republic Revisited}, pages
  66--105. Oxford University Press, Oxford, 2007.

\bibitem[Zanardi et~al.(2004)Zanardi, Lidar, and
  Lloyd]{2004_zanardi_lidar_lloyd}
Paolo Zanardi, Daniel~A Lidar, and Seth Lloyd.
\newblock Quantum tensor product structures are observable induced.
\newblock \emph{Physical review letters}, 92\penalty0 (6):\penalty0 060402,
  2004.

\bibitem[Zurek(1989)]{1989_zurek}
Wojciech~H Zurek.
\newblock Algorithmic randomness and physical entropy.
\newblock \emph{Physical Review A}, 40\penalty0 (8):\penalty0 4731, 1989.

\bibitem[{\.Z}yczkowski et~al.(1998){\.Z}yczkowski, Horodecki, Sanpera, and
  Lewenstein]{1998_zyczkowski_horodecki_sanpera_lewenstein}
Karol {\.Z}yczkowski, Pawe{\l} Horodecki, Anna Sanpera, and Maciej Lewenstein.
\newblock Volume of the set of separable states.
\newblock \emph{Physical Review A}, 58\penalty0 (2):\penalty0 883, 1998.

\end{thebibliography}

\end{document}